\documentclass[aps,prb,twocolumn,longbibliography,superscriptaddress]{revtex4-2}

\usepackage{tabularx}   
\usepackage{booktabs}   
\usepackage{array}      

\usepackage{ulem}
\usepackage{amsmath,amssymb,bm,graphicx}
\usepackage{dcolumn}
\usepackage{color}
\usepackage{xcolor}  
\usepackage{soul} 
\usepackage[utf8]{inputenc}
\usepackage[T1]{fontenc}
\usepackage[colorlinks=true,citecolor=blue,linkcolor=blue,urlcolor=blue]{hyperref}
\definecolor{goldenrod}{rgb}{0.85, 0.65, 0.13}

\begin{document}

\title{Vortex structure and intervortex interaction in superconducting structures with intrinsic diode effect}

\author{A.~V.~Putilov}
\email{alputilov@ipmras.ru}
\affiliation{Moscow Institute of Physics and Technology, Dolgoprudnyi, Moscow Region 141701, Russia}
\affiliation{Institute for Physics of Microstructures, Russian Academy of Sciences, 603950 Nizhny Novgorod, GSP-105, Russia}

\author{D.~V.~Zakharov}
\affiliation{ITMO University, Saint-Petersburg, Russia}

\author{A.~Kudlis}
\affiliation{Science Institute, University of Iceland, Dunhagi 3, IS-107 Reykjavik, Iceland}

\author{A.~S.~Mel’nikov}
\affiliation{Moscow Institute of Physics and Technology, Dolgoprudnyi, Moscow Region 141701, Russia}
\affiliation{Institute for Physics of Microstructures, Russian Academy of Sciences, 603950 Nizhny Novgorod, GSP-105, Russia}

\author{A.~I.~Buzdin}
\affiliation{University of Bordeaux, LOMA UMR-CNRS 5798, F-33405 Talence Cedex, France}
\affiliation{Institute for Computer Science and Mathematical Modeling, Sechenov First Moscow State Medical University, Moscow, 19991, Russia}
\date{\today}

\begin{abstract}
We demonstrate that the intrinsic superconducting diode effect can affect the structure, interactions and dynamics of Abrikosov vortices in non-centrosymmetric superconductor/ferromagnet hybrid structures.
The Ginzburg-Landau (GL) theory accounting for the spin-orbit and exchange-field effects predicts a chiral distortion of the superfluid velocity, non-central interaction forces and resulting torque in a vortex-antivortex pair, and anisotropy of the Bean-Livingston barrier. These closed-form results are fully confirmed by time-dependent GL numerical simulations carried out with a fourth-order least-squares finite-difference solver, which captures equilibrium single vortex configuration in realistic mesoscopic geometries.  The analysis shows that the cubic gradient term shifts vortex cores by an amount proportional to the in-plane exchange field and simultaneously generates a lateral torque that can rotate entire vortex ensembles, showing how spin-orbit coupling (SOC) and the exchange field enable breakdown of the vortex--antivortex symmetry in a finite-size sample.
By combining transparent analytics with quantitative numerics, the work provides the hallmarks of vortex physics in superconducting structures with an intrinsic diode effect and supplies concrete guidelines for designing non-reciprocal superconducting circuits, fluxonic logic elements, and kinetic-inductance devices.
\end{abstract}

\maketitle

\section{Introduction}

For several decades hybrid nanostructures that combine superconducting (S) and ferromagnetic (F) components have offered a versatile playground for probing the interplay between pairing correlations and magnetism \cite{Buzdin-RMP-05, Mironov-UFN-22,Bergeret-RMP-05}.  Besides the conventional proximity coupling of the two subsystems, the breaking of inversion symmetry at the S/F interface and the presence of a sizeable spin-orbit interaction (SOI) unlock a number of non-trivial phenomena. 
\begin{figure}[t!]
    \centering
    \includegraphics[width=0.9\linewidth]{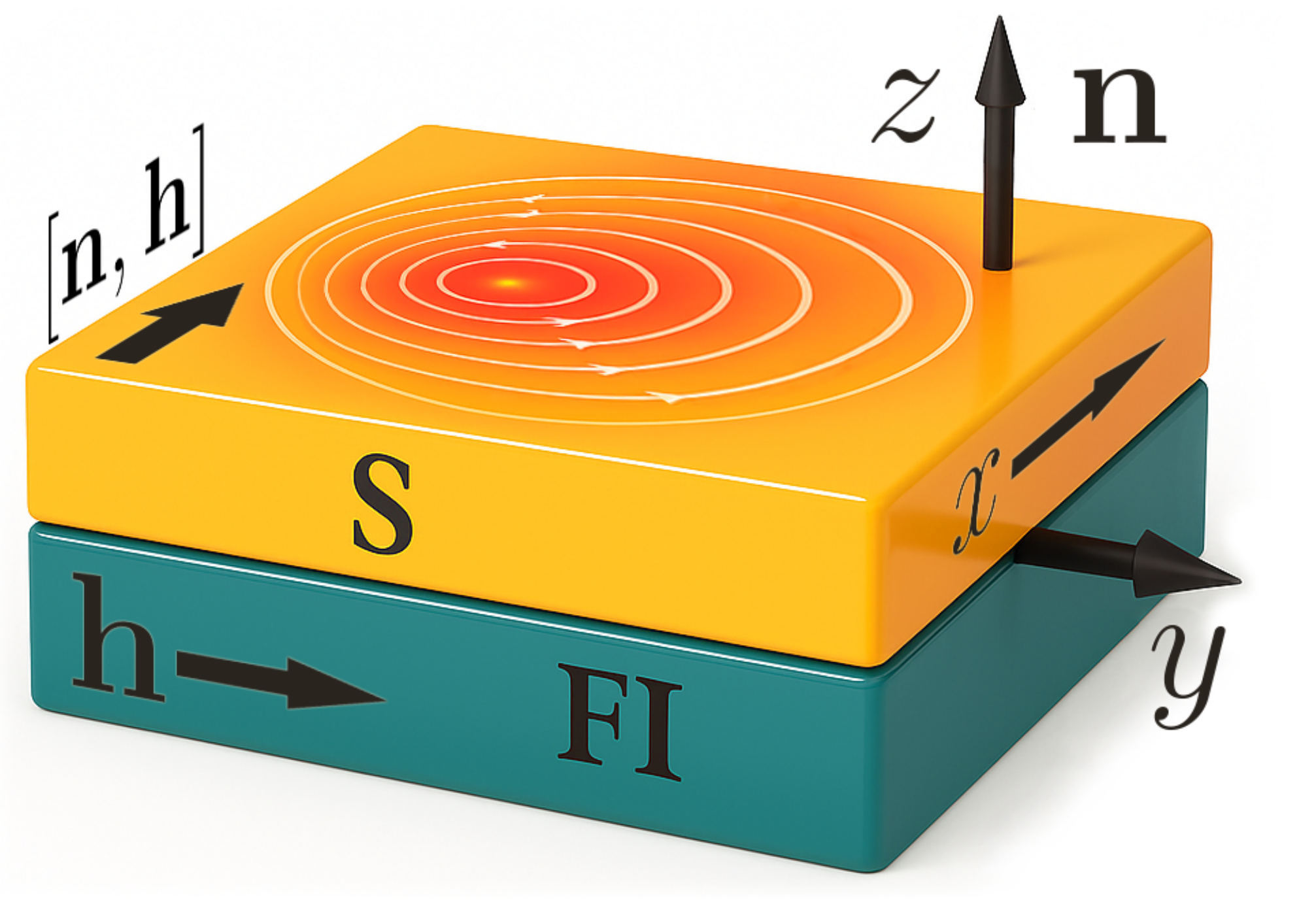}
    \caption{A thin superconducting layer (orange, S) rests on a ferromagnetic insulator (teal, FI).  The ferromagnet’s uniform in‐plane exchange field $\mathbf{h}$ and the interface normal $\mathbf{n}$ (along $z$) produce the intrinsic direction $[\mathbf{n}\times\mathbf{h}]$ in the $xy$-plane.  In the superconductor we depict the circular supercurrent of a single vortex, whose profile is deformed by the cubic spin-orbit term aligned with $x$.  The axes $x$, $y$, and $z$ indicate the laboratory frame; arrows show the directions of $\mathbf{h}$, $\mathbf{n}$, and the vortex circulation.}
    \label{fig:skt}
\end{figure}
Prominent examples include Majorana edge modes in topological S/F hybrids~\cite{Alicea}, Josephson $\varphi_{0}$ junctions in spin-polarized weak links~\cite{Buzdin_Phi,Krive,Reynoso,Kouwenhoven}, and helical ground states supporting spontaneous supercurrents~\cite{Mironov-PRL-17,Devizorova-PRB-21}. 

A particularly striking manifestation of this magnetochiral anisotropy is the \textit{superconducting diode effect} -- nonreciprocal critical currents and current--voltage characteristics in resistive state.
First observed in spatially inhomogeneous Josephson junctions and superconducting ratchets \cite{Krasnov-PRB-97,Villegas-Sci-03} this diode effect has been further analyzed for a great variety of systems with different mechanisms of magnetochiral anisotropy \cite{Silaev-JPCM-14,Lyu-NatComm-21,Majer-PRL-03,Vodolazov-PRB-05,deSilva-NatLett-06,Suri-APL-22}.
More recently the nonreciprocal phenomena 
have been observed in
multilayers lacking spatial inversion symmetry~\cite{Ando-Nature-21,Bauriedl-NatComm-22}, ballistic Josephson devices~\cite{Baumgartner-NatNano-22} and in S/F multilayers with an in-plane exchange field~\cite{Narita-NatNano-22}.
Different theoretical scenarios describing the intrinsic diode effect have been suggested and analyzed in the context of existing and proposed experiments \cite{Silaev-JPCM-14,Fominov-PRB-22,Daido-PRL-22,Ilic-PRL-22,Devizorova-PRB-21,Putilov-PRB-24,Kochan-Arxiv-2023}.
Focusing on the particular case of the nonreciprocal effects associated with the spin-orbit and exchange field one can formulate a rather general description of the intrinsic diode effect based on modification of the standard Ginzburg--Landau (GL) type theory. The required symmetry breaking can be described by the additional terms odd in the spatial gradients of the superconducting order parameter \cite{Mineev-JETP-94,Edelstein-JPCM-96,Kaur-PRL-05}. Accounting the linear and cubic gradient terms allows to describe the asymmetry in critical currents \cite{Daido-PRL-22,He-NJP-22}, appearance of spontaneous currents in nonuniform systems of different geometry \cite{Mironov-PRL-17}, photogalvanic phenomena \cite{Mironov-PRB-24}, and direct coupling between supercurrent and magnetic moment \cite{Plastovets-PRB-24}.

One can naturally expect that
the intrinsic diode effect should cause drastic modifications and provoke intriguing new phenomena in the properties of vortex matter, including structure of individual vortices and vortex lattices as well as the vortex dynamics.
Recent experiments already hint at such modifications. In a Rashba-coupled meander microstrip the vortex-induced kinetic inductance was shown to depend on the direction of an in-plane magnetic field~\cite{Fuchs-PRX-22}.  
Thus, it seems timely to develop a theoretical model that describes possible non-reciprocal phenomena in the vortex state.

In the present work, we make a first step to close this gap considering the structure of individual vortices and vortex interactions in the low magnetic field regime, i.e., for low vortex concentrations.  As a specific platform, we consider a thin superconducting film deposited on an \emph{insulating} ferromagnet. In Fig.~\ref{fig:skt}, we schematically present the analyzed system. The interface normal $\mathbf n$ breaks inversion symmetry, while an in-plane exchange field $\mathbf h$ breaks time-reversal symmetry; taking $\mathbf h$ parallel to the layers suppresses stray fields. Because the ferromagnet is insulating, proximity suppression of superconductivity is negligible. The thickness $d$ of the superconducting film must be relatively small to make surface effects more pronounced, and the vectors $\mathbf n$ and $\mathbf h$ together single out the in-plane direction $[\mathbf n\times\mathbf h]$ that governs all non-reciprocal responses discussed below. 
Here we neglect the effects of Zeeman field, which can also lead to distortion of the vortex magnetic field in crystals with broken inversion symmetry
\cite{Agterberg-PRB-07,Lu-PRB-08,Lu-JLTP-09,Yip-JLTP-2005}.
Spin-orbit coupling at the superconductor/ferromagnet interface enters the GL theory through odd spatial-derivative terms whose strength is set by two phenomenological constants. Averaging these contributions across a film only a coherence-length thick produces an effective two-dimensional free-energy functional that augments the usual GL terms as well as the linear and cubic gradient contributions responsible for the intrinsic diode effect. The linear spin-orbit term forces the equilibrium condensate to form a helical pattern whose wave vector points along the cross-product of the interface normal and the in-plane exchange field; the cubic gradient term gives the asymmetry in the nonlinear relation between the current density and superfluid velocity giving, thus, the nonreciprocal response. 

Below, analyzing this setup, we (i) derive analytical expressions for the chiral distortion of the superfluid velocity around a single vortex, (ii) calculate the  non-central force in a vortex-antivortex pair, and (iii) obtain the direction-dependent correction to the Bean-Livingston barrier in different geometries.  All predictions are verified by time-dependent GL simulations that employ a fourth-order least-squares finite-difference scheme and quantitatively reproduce the analytic results.

The paper is organized as follows. Section \ref{sec:model} formulates the two-dimensional GL functional with interfacial spin-orbit terms and reduces it to a working model for analytical and numerical simulations. In Section \ref{sec:single_vortex} we determine the supercurrent distribution around an isolated vortex in the presence of cubic gradient terms in the GL functional. Section~\ref{sec:interaction_vort} evaluates the corresponding free energy and extracts the non-central force acting inside a vortex-antivortex pair, Section~\ref{sec:vort_disk} extends the analysis to a single vortex in a superconducting disk, while Section~\ref{sec:vort_boundary} establishes the modification of the Bean-Livingston barrier for a vortex near a plane boundary.
Section~\ref{sec:motion} describes the peculiarities of the vortex dynamics.
Section~\ref{sec:numerical} compares all analytical predictions with time-dependent GL simulations that employ a high-accuracy least-squares finite-difference solver. Technical details of the numerical scheme and supplementary derivations are collected in the Appendix.

\section{Model}\label{sec:model}

In this section we formulate the GL model aimed to describe the vortex physics in the thin film geometry shown in Fig.\ref{fig:skt}. 
As it has been noted in introduction the magnetochiral anisotropy triggered by the exchange field and the interfacial spin-orbit interaction can be incorporated into the GL free-energy density through the odd-gradient terms \cite{Mineev-JETP-94, Edelstein-JPCM-96, Kaur-PRL-05, Mironov-PRL-17, Daido-PRL-22}.  
Since the SOI is confined to an atomic-scale interface layer \cite{Mironov-PRL-17} and the film thickness $d$ is of the order of coherence length $\xi$, variations of the order parameter along $z$ can be neglected; averaging over $d$ then leads to the two-dimensional density of free energy
\begin{align}
F=& -\alpha|\Psi|^{2}
          +\tfrac{\beta}{2}|\Psi|^{4}
          +\xi^{2}\alpha|{\bf D}\Psi|^{2}\label{Eq_GLfunc}  \\
          &+\bigl(\varepsilon_{1}\Psi^{*}[\mathbf n\!\times\!\mathbf h]\mathbf D\Psi
          +\varepsilon_{3}({\bf D}\Psi)^{*}[\mathbf n\!\times\!\mathbf h]\mathbf D^{2}\Psi
          +\text{c.c.}\bigr),\nonumber 
\end{align}
where $\alpha=\alpha_{0}(1-T/T_{c0})$, $\beta>0$, $T_{c0}$ is the superconducting critical temperature and operator $\mathbf D=-i\nabla-2\pi\mathbf A/\Phi_0$ where $\mathbf A$ is a vector potential.  
The phenomenological coefficients $\varepsilon_{1}$ and $\varepsilon_{3}$ quantify the strength of the linear and cubic gradient terms, respectively. The higher order odd and even gradient terms are omitted as we consider large scale spatial variations of the order parameter in the plane of the 
superconducting layer.
Spin-orbit coupling together with the exchange interaction fixes a helical ground state whose order parameter takes the form $\Psi=|\Psi|e^{i\mathbf{qx}}$, with the pitch vector $\mathbf q$ aligned along $[\mathbf n\times\mathbf h]$\,\cite{Edelstein-JPCM-96,Mineev-JETP-94}.  


Choosing the $x$ axis along the intrinsic direction
$[\mathbf n\times\mathbf h]$, variation of the functional \eqref{Eq_GLfunc} gives
\begin{multline}
\alpha\Psi-\beta|\Psi|^{2}\Psi-\xi^{2}\alpha\mathbf D^{2}\Psi \\
+2\varepsilon_{1}h\,\mathbf D_{x}\Psi
+2\varepsilon_{3}h\,\mathbf D_{x}\mathbf D^{2}\Psi=0 .
\label{Eq_GLeq}
\end{multline}
To remove the linear SOI term we write $\Psi=\Psi_0\,\psi\,e^{iqx}$ with 
$q=\left(\alpha\xi^2-\sqrt{\alpha^2\xi^4-6h^2\varepsilon_1\varepsilon_3}\right)/(3h\varepsilon_3)$ and $\Psi_0=\sqrt{\alpha'/\beta}$, where $\alpha'=\alpha+2\varepsilon_{1}hq-\xi^{2}\alpha q^{2}-2\varepsilon_{3}q^{3}$.
Finally we obtain
\begin{multline}
\alpha'\psi-\alpha'|\psi|^{2}\psi
+(\xi^{2}\alpha+2\varepsilon_{3}hq)\mathbf D^{2}\psi
+4q\varepsilon_{3}h\,\mathbf D_{x}^{2}\psi \\
-2\varepsilon_{3}h\,\mathbf D_{x}\mathbf D^2\psi = 0.
\label{Eq_GLeq1}
\end{multline}
Leaving only effects linear in parameters $\varepsilon_1h$ and $\varepsilon_3h$ describing the magnito-chiral anisotropy we get approximate expressions for $q=\varepsilon_{1}h/(\alpha\xi^{2})$, $\alpha'=\alpha$. In most part of the paper (except the Sec.V) we neglect the terms proportional to $hq\varepsilon_3$ omitting, thus, anisotropic correction to the effective mass tensor.
In principle, this anisotropy can be excluded by scaling transformation of coordinates.
Thus, we obtain a basic GL equation in the following form:
\begin{equation}
\psi-|\psi|^{2}\psi-\xi^2\mathbf D^{2}\psi
-2\gamma\,\xi^3\mathbf D_{x}{\mathbf D}^{2}\psi=0 , 
\label{Eq_GLfin}
\end{equation}
where the single cubic-gradient dimensionless parameter
$\gamma=\varepsilon_{3}h/(\alpha\xi^{3})$ encapsulates the strength of the
interfacial spin-orbit coupling and exchange field.
We use natural boundary conditions arising from integration by parts in the standard method of varying the functional with respect to the order parameter:
\begin{align}
\textbf{N} \cdot i\gamma{\mathbf D}\psi|_{S} = 0; \nonumber \\ 
\textbf{N} \cdot\Big( {\mathbf D} + \gamma\xi {\mathbf D}{\mathbf D}_x  + \gamma\xi\textbf{x}_{0} {\mathbf D}^2 \Big) \psi|_{S} = 0,
 \end{align}
where $\mathbf x_0$ is the unit vector along $x$ axis and $\mathbf N$ is an in-plane unit vector normal to the film edge.

Varying the GL functional with respect to the vector potential we get the Maxwell equations with the currents flowing only inside a very thin superconducting layer of thickness $d$. These equations contain a natural length scale of magnetic screening $\lambda_{eff}=\lambda^2/d$, where $\lambda$ is the London penetration depths of the superconducting material \cite{Pearl-APL-64}.
This effective screening length should strongly exceed the coherence length defining the size of vortex cores and can even exceed the lateral size of the sample. In the latter case, we can ignore these screening effects at all and take the vector potential determined only by the external applied magnetic field.
This limit vanishing screening will be exploited, in particular, in our numerical simulations in section \ref{sec:numerical}.

\section{Single isolated vortex within the London approximation}\label{sec:single_vortex}


In this section we consider an isolated vortex created by a weak perpendicular field. 
For large distances $r\gg\xi$ from the vortex center, i. e., beyond the vortex core, we neglect the variation of absolute value of the order parameter $|\psi|$ and use, thus, the London approach.
In this limit the free energy density takes the form
\begin{equation}
F=\frac{H_c^2}{4\pi}|\psi|^2\!\left[-1+\frac{|\psi|^2}{2}
        +\xi^2\bigl(\mathbf v^{2}+2\gamma\xi(\mathbf x_0\mathbf v)\mathbf v^{2}\bigr)\right],
\end{equation}
where $\psi=|\psi|e^{i\varphi}$, $H_c^{2}=4\pi\alpha^{2}/\beta$, 
and the superfluid velocity $\mathbf v=\nabla\varphi-2\pi\mathbf A/\Phi_{0}$.
The associated current density is
\begin{equation}
\mathbf j=-c\frac{\partial F}{\partial\mathbf A}
         =\frac{c\Phi_{0}}{8\pi^{2}\lambda^{2}}|\psi|^{2}
         \Bigl[\mathbf v+2\gamma\xi(\mathbf x_{0}\mathbf v)\mathbf v+\gamma\xi\mathbf v^{2}\mathbf x_{0}\Bigr].
\end{equation}
At distances $r\ll\lambda_{eff}$ from the vortex center, the screening effects are negligible and the vector potential $A$ can be omitted in the definition of $\mathbf v$.
Within the perturbative expansion in a small parameter $\gamma$, we write 
$\varphi=\varphi_{0}+\varphi_{1}$ and $\mathbf v=\mathbf v_{0}+\mathbf v_{1}$.  
For $\gamma=0$ and at distances $r\gg\xi$ the unperturbed vortex is described by the order parameter 
$\psi_{0}(r,\theta)=e^{i\theta}$ and corresponding superfluid velocity $\mathbf v_{0}=\pmb\theta_{0}/r$, giving us
\begin{equation}
\mathbf j=\frac{c\Phi_{0}}{8\pi^{2}\lambda^{2}}|\psi|^{2}
          \left(\frac{\pmb\theta_{0}}{r}
                +\mathbf v_{1}
                -2\pmb\theta_{0}\frac{\gamma\xi}{r^{2}}\sin\theta
                +\gamma\xi\frac{\mathbf x_0}{r^{2}}\right),
\label{Eq_Curr1}
\end{equation}
where charge conservation $\nabla\!\cdot\!\mathbf j=0$ implies 
\begin{align}
\varphi_{1,rr}+r^{-1}\varphi_{1,r}+r^{-2}\varphi_{1,\theta\theta}
            =4\gamma\xi r^{-3}\cos\theta.
\end{align}
Its particular solution that vanishes at infinity looks as follows:
\begin{align}
\varphi_{1}=-2\gamma\cos\theta\,\frac{\ln \tilde r}{\tilde r},
\end{align}
where $\tilde r=r/\xi$ is a dimensionless coordinate. Hence, far from the core ($\tilde r\gg1$) the order parameter takes the form
\begin{align}
\!\psi=\!e^{i\theta}\!\exp\!\left(i\varphi_1\right)
     \!\approx\! e^{i\theta}\!\!\left(\!1\!-\!2i\gamma\frac{\ln\tilde r}{\tilde r}\cos\theta\!\right).\!\!
\end{align}
Substituting this result into Eq.~\eqref{Eq_Curr1} yields the current distribution
\begin{align}
\mathbf j=\frac{c\Phi_{0}}{8\pi^{2}\lambda^{2}\xi}
          \left(\frac{\pmb\theta_{0}}{\tilde r}+\gamma\mathbf J\right),
\end{align}
with
\begin{align}\label{j_teor}
\mathbf J&=\frac{1}{\tilde r^{2}}
           \Bigl[\pmb r_{0}\cos\theta\,(2\ln\tilde r-1)
                 +\pmb\theta_{0}\sin\theta\,(2\ln\tilde r-3)\Bigr] \\
         &=\frac{\mathbf x_{0}}{\tilde r^{2}}
           \bigl[2\cos2\theta(\ln\tilde r-1)+1\bigr]
           +\frac{\mathbf y_{0}}{\tilde r^{2}}\,2\sin2\theta(\ln \tilde r-1).\nonumber
\end{align}
At logarithmically large distances ($\ln\tilde r\gg1$) this simplifies to
\begin{align}
\mathbf J\simeq\frac{2\ln\tilde r}{\tilde r^{2}}
               \bigl(\pmb r_{0}\cos\theta+\pmb\theta_{0}\sin\theta\bigr).
\end{align}

All subsequent sections are built on these expressions to quantify vortex interactions and edge effects in the presence of intrinsic diode effect.

\section{Interaction of vortices}
\begin{figure}[t]
\centering
\includegraphics[width=.85\linewidth]{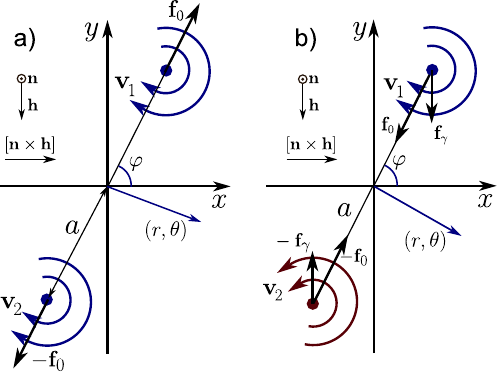}
\caption{Schematic geometry used to analyze the intervortex interaction. We introduce here a polar coordinate system ${\bf r}=(r,\theta)$, the coordinate origin is placed midway between the two vortices. a) Two vortices with coinciding vorticity which generates superfluid velocities $\mathbf v_{1}$ and $\mathbf v_2$. b) Analogous configuration of two vortices with opposite vorticities. The conventional intervortex interaction forces $\mathbf f_0$ and $\gamma$-dependent non-central forces $\mathbf f_\gamma$ are shown.}
\label{Fig_V_AV}
\end{figure}
We now turn to analysis of modification of vortex interaction due to the cubic gradient term.
Starting from the full GL free energy, we examine both the mutual interaction of two vortices (with coinciding or opposite vorticity) and the interaction of a single vortex with a sample boundary. For a vortex pair the key quantity is the total free energy $U(r_{1},r_{2})$ as a function of their coordinates; the force on vortex~$i$ is defined through the expression
\begin{equation}
\mathbf f_i=-\frac{\partial U(\mathbf r_1,\mathbf r_2)}{\partial \mathbf r_i},
\end{equation}
with $i=1,2$.  
Treating the cubic spin-orbit term as a perturbation, we write the free energy up to the first order in $\gamma$ as
\begin{equation}
U[\mathbf v]= \int \Bigl[F_0+\frac{\partial F_0}{\partial\mathbf v}\mathbf v_1+\frac{H_c^2\xi^3}{2\pi}|\psi|^{2}\gamma(\mathbf x_0,\mathbf v_0)\mathbf v_0^{2}\Bigr]dV ,
\end{equation}
where $F_{0}$ is the conventional GL energy density (in the case $\gamma=0$) and $\mathbf v_0$ is the unperturbed superfluid velocity.
Substituting the unperturbed solution $\mathbf v_0$ of the GL equation into the above functional we get the zero first variation $\partial F/\partial\mathbf v=0$. 
Thus, the integral can be evaluated without calculation of the first‐order correction:
\begin{equation}\label{Eq_FE}
U=U_0+\frac{H_c^2\xi^3}{2\pi}\!
      \int |\psi|^{2}\gamma(\mathbf x_0,\mathbf v_0)\mathbf v_0^{2}dV,
\end{equation}
The expression is valid as far as the characteristic lateral dimensions $a$ satisfies the condition $\xi\ll a\ll\lambda_{eff}$.

\subsection{Two vortices in an infinite film \label{sec:interaction_vort}}
We continue with the calculation of the interaction energy (\ref{Eq_FE}) for a pair of vortices in an infinite film.
The vortex centers are taken at the points $(a,\varphi)$ and $(a,\varphi+\pi)$ in polar coordinates, see Fig.~\ref{Fig_V_AV}.  
A velocity field $\mathbf v=\pmb\theta/r$ corresponds to the vorticity $+1$, while the reversed flow $\mathbf v=-\pmb\theta/r$ has the vorticity $-1$.  
When two vorticities have the \textit{same} sign (Fig.~\ref{Fig_V_AV}a), the integrand in Eq.~\eqref{Eq_FE} is odd, $\mathbf v_{0}(\mathbf r)=-\mathbf v_{0}(-\mathbf r)$, so the contribution linear in the parameter~$\gamma$ cancels after integration.  
Thus, the first-order $\gamma$-correction to the vortex-vortex interaction vanishes.

The leading $\gamma$-correction  appears only for a vortex-antivortex pair (Fig.~\ref{Fig_V_AV}b), for which $\mathbf v_{0}(\mathbf r)=\mathbf v_{0}(-\mathbf r)$.  
We start calculation of the interaction energy (\ref{Eq_FE}) for the case $\varphi=\pi/2$, for which 
the unperturbed velocity field produced by the two vortices with opposite vorticity is  
\begin{equation}
\mathbf v_{1,2}= \frac{-\mathbf x_{0}(a\mp r\sin\theta)-\mathbf y_{0}\cos\theta}
                       {(a\mp r\sin\theta)^{2}+r^{2}\cos^{2}\theta}.
\end{equation}
A straightforward algebra gives
\begin{equation}
(\mathbf v_1+\mathbf v_2)^{2}
  (\mathbf v_{1x}+\mathbf v_{2x})
  \!=\!-\frac{8a^{3}(a^{2}\!+\!r^{2}\!\cos\!2\theta)}
         {(a^{4}\!+\!r^{4}\!+\!2a^{2}r^{2}\!\cos\!2\theta)^{2}} .
\end{equation}
Substituting into the Eq.~\eqref{Eq_FE} we obtain
\begin{equation}
U=U_0-\frac{dH_c^2\xi^3\gamma_x}{2\pi}
        \int(\mathbf v_1+\mathbf v_2)^{2}
             (\mathbf v_{1x}+\mathbf v_{2x})\,r\,dr\,d\theta .
\end{equation}
Performing the angular and radial integration
and cutting the divergence at the core radius $r=\xi$, we find
\begin{equation}
U\simeq U_0+\frac{2dH_c^2\xi^{3}\gamma}{a}\ln\frac{\xi}{a}.
\end{equation}
The result can be easily generalized for arbitrary arrangement of vortices. One can write
\begin{equation}
U\simeq U_0+\sin\varphi\,
        \frac{2dH_c^2\xi^{3}\gamma}{a}\ln\frac{\xi}{a},
\end{equation}
with $\varphi$ the angle between the line joining the cores and the intrinsic direction $[\mathbf n\times\mathbf h]$.
The resulting force is
\begin{equation}
\mathbf f=-\nabla U
         =\mathbf f_{0}+4dH_c^2\xi^{3}\gamma\,
            \nabla\!\left[\frac{\sin\varphi\,\ln r}{r}\right],
\end{equation}
or explicitly, in the limit $\ln(a/\xi)\gg1$
\begin{equation}
\mathbf f=\mathbf f_{0}
          -4\mathbf y_0dH_c^2\xi^{3}\gamma\frac{\ln a}{a^{2}}          
          =\mathbf f_{0}+\mathbf f_\gamma,
\end{equation}
where $\mathbf f_{0}$ is the conventional attractive force, and $\mathbf f_\gamma$ 
is $\gamma$-dependent new \textit{non-central} force that tends to rotate the pair about their midpoint.

\subsection{Vortex in a superconducting disk \label{sec:vort_disk}}

\begin{figure}[t!]
\includegraphics[width=0.6\linewidth]{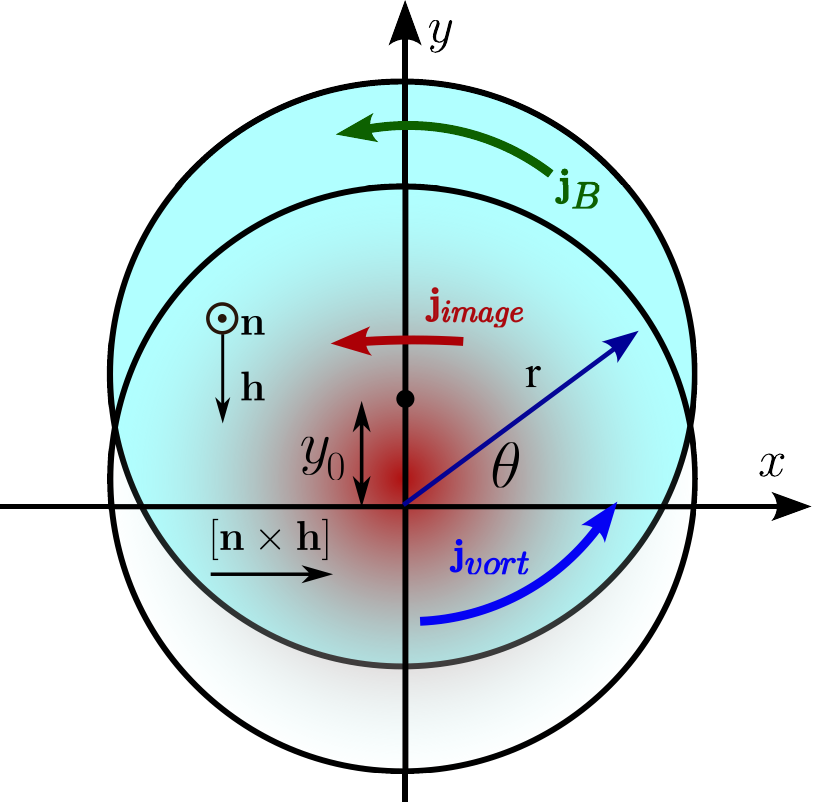}
\caption{A schematic view of a single vortex confined in a superconducting disk of radius $R$. A center of polar coordinate system $(r,\theta)$ is placed in the vortex center and shifted from the disk center by the distance $y_0$. A vortex-induced current $\mathbf j_{vort}$, screening current $\mathbf j_B$ and current of an image vortex $\mathbf j_{image}$ are shown by blue, green and red, respectively.
}
\label{Fig_Disk}
\end{figure}


We now turn to the study of the interaction of a single vortex with sample boundaries.  
 According to the textbook picture the sample edge affects the vortex through the following mechanisms: (i) Lorentz force caused by the Meissner screening currents flowing along the boundary pushes the vortex into the sample; (ii) the boundary condition of zero current flowing through the edge results in the appearance of the force attracting the vortex to the boundary which can be viewed as the force acting between the vortex and the image antivortex placed outside the superconducting region.

As a first and instructive example we consider a vortex in a mesoscopic superconducting disc of the radius $R$. When the parameter $\gamma$ is set to zero the energy of a vortex whose core is displaced by a distance $y$ from the disc center takes the form~\cite{Buzdin-PLA-94}
\begin{equation}
U
 =\frac{dH_{c}^{2}\xi^{2}}{2}
   \Bigl[
      \ln\frac{R}{\xi}
     +\ln\!\Bigl(1-\frac{y^{2}}{R^{2}}\Bigr)
     -b\!\Bigl(1-\frac{y^{2}}{R^{2}}\Bigr)
     +\frac{b^{2}}{4}
   \Bigr],
\label{Eq_B}
\end{equation}
where the dimensionless field $b=\pi R^{2}B/\Phi_{0}$ equals to the number of flux quanta through the disc, here $B$ is an external magnetic field.  
For a vortex to be trapped inside the disc one needs $b>1$, i.e. at least one flux quantum piercing the sample.

The nonreciprocal contribution to the energy can be evaluated from the expression Eq.~\eqref{Eq_FE}.  
We choose polar coordinates $(r,\theta)$ whose origin coincides with the
vortex center displaced by the distance $y$ from the disc center
(see Fig.~\ref{Fig_Disk}). In these coordinates the unperturbed super-velocity reads  
\begin{align}\label{eq:v_disk}
v_{x}&=\frac{\sin\theta}{r}
       -\frac{b}{R^2}\left(r\sin\theta-y\right)
       -\frac{y}{R^{2}},\nonumber\\
v_{y}&=-\frac{\cos\theta}{r}
       +\frac{b}{R^2}r\cos\theta,
\end{align}
where the first term is produced by the vortex itself, the second term comes from the
screening currents generated by the external field $B$, and the third term originates from the interaction with the image
antivortex situated at the distance $R^{2}/y$ along the negative $y$
direction (in the latter term only contributions linear in $y$ are retained).

Adding the $\gamma$-dependent correction from Eq.~\eqref{Eq_FE} to the
field-induced energy~\eqref{Eq_B} and expanding the resulting expression for $y\ll R$ we obtain  
\begin{multline}
U = \frac{dH_{c}^{2}\xi^{2}}{2}
    \Bigl[
        \ln\frac{R}{\xi}
       -b
       -\frac{y^{2}}{R^{2}}(1-b)+\frac{b^2}{4}\\
       {}-\;
       y\,\frac{\gamma\xi(1-b)}{R^{2}}
       \bigl(-4\ln\tfrac{R}{\xi}+1+2b\bigr)
    \Bigr].
\end{multline}
Minimizing $U$ with respect to $y$ yields the equilibrium displacement  
\begin{equation}
y_0= \gamma\xi\bigl[-2\ln(R/\xi)+\pi R^2B/\Phi_0+0.5\bigr].
\label{shift teor}
\end{equation}
The sign of the vortex shift depends both on the sign of the parameter $\gamma$ (i.e., on the direction of the exchange field $\mathbf{h}$ and the vector $\mathbf{n}$ entering the spin-orbit interaction) and on the sign of multiplier $0.5+\pi R^2B/\Phi_0-2\ln(R/\xi)$, so the shift direction can be
reversed by tuning the perpendicular magnetic field.
Note that for a large disk with $R\ge\lambda_{eff}$ the screening currents become important suppressing the vortex-edge interaction and the related shift $y_0$ with further increase of $R$.
\begin{figure}[t!]
\includegraphics[width=0.6\linewidth]{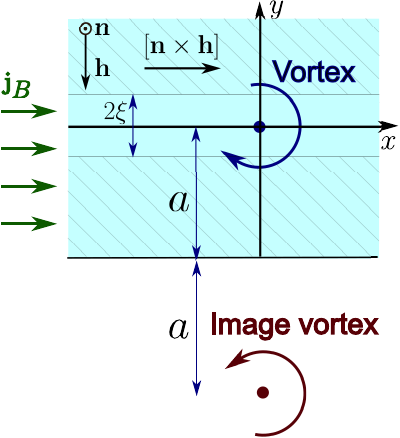}
\caption{Schematic picture of a single vortex located at the distance $a$ from the superconducting film edge.  Curved arrows show the direction of the supercurrents around the real (blue) vortex and image (red) antivortex. The area of integration is indicated by shading.}\label{fig:boundary}
\end{figure}

\subsection{Vortex near a plane boundary \label{sec:vort_boundary}}

Now we proceed with consideration of the vortex interaction with a plane boundary.
In the same way as we did in the case of the disk, we take the superfluid velocity as a superposition of the contribution of the vortex itself, which is situated at a distance $a$ from the film edge, the vortex image with an opposite vorticity, and the current density $\mathbf j$ flowing near the superconducting film edge. The latter contribution appears either in the presence of an applied transport current or an external magnetic field.
The calculation of the relation between the current density $\mathbf j=\mathbf x_0j_x$ and the external magnetic field requires the solution of the Maxwell equations taking into account  the sample geometry.
For the Cartesian coordinate system with a center coinciding with a vortex center (Fig. \ref{fig:boundary}) this velocity takes the form
\begin{align}
v_{x}&=\frac{y}{x^2+y^2}+\frac{8\pi^2\lambda^2j_x}{\Phi_0}
       -\frac{2a+y}{x^2+(2a+y)^2},\nonumber\\
v_{y}&=-\frac{x}{x^2+y^2}
       +\frac{x}{x^2+(2a+y)^2}.
\end{align} 
To find the interaction force, we calculate the system free energy using the expression (\ref{Eq_FE}).
The resulting integral diverges logarithmically at the origin of the coordinate system and must be regularized by introducing a lower cutoff at the core size $\xi$.
Therefore, we perform integration over the $x$ axis in infinite limits and subsequent integration over the $y$ axis within the limits $y\in(-a,-\xi)\cup(\xi,+\infty)$, the integration area is indicated in Fig. \ref{fig:boundary} by shading.

The total free energy of a vortex at the distance $a$ from the edge then becomes
\begin{multline}\label{eq:BLbarrier}
U
=\frac{\Phi_{0}d}{c}\Bigl[j_xa-j_0\xi\ln\frac{a}{\xi}
\;+\;\gamma j_x\xi\Bigl(1-\frac{3j_xa}{2j_0\xi}\Bigr)\\
-2\gamma j_0\xi^2
\Bigl(\frac{1}{a}-\frac{j_x}{j_0\xi}\Bigr)\ln\frac{a}{\xi}
\Bigr],
\end{multline}
where $j_0=c\Phi_0/(16\pi^2\xi\lambda^2)$ 
is of the order of the depairing current \cite{Tinkham-book-96}.
In the Eq. (\ref{eq:BLbarrier}) the first term describes the work of the current $\mathbf j$, the second term is energy of interaction with the image antivortex, and the $\gamma$-dependent terms arise from the cubic diode term in the GL functional.
The maximum value of the $U(a)$ function defines the height of the Bean-Livingston barrier for the vortex entry.  For $\gamma=0$, $U(a)$ reaches its maximum at $a\sim\xi j_0/j_x$. The cubic-gradient term produces the correction to the barrier height
\begin{equation}\label{bin correction}
\delta U=\pm\frac{\gamma\Phi_0d\xi}{2c}j_x+\mathcal{O}(\gamma^2).
\end{equation}
The sign of the value $\delta U$ 
is determined by the signs of the spin--orbit interaction constant $\varepsilon_3$ and of the mixed product $[\mathbf n\times\mathbf h]\cdot\mathbf j$.

Along with the change in the barrier height, the minimal current $j$ required to overcome this barrier also changes. To calculate this critical current of the vortex entry, we find the force acting on the vortex near the edge $f=-\partial U/\partial a$ from the expression (\ref{eq:BLbarrier}):
\begin{equation}
\frac{\partial U}{\partial a}
\propto \frac{j_x}{j_0}-\frac{\xi}{a}-
\gamma \left[\frac{3j_x^2}{2j_0^2}+
\frac{2\xi^2}{a^2}\left(\ln\frac{a}{\xi}-1\right)
+\frac{2\xi j_x}{aj_0}\right],
\end{equation}
The vortex enters the sample through the barrier when the top of the barrier is at a distance of the order of $\xi$ from the edge of the sample.
Substituting $a\sim\xi$ and assuming $\gamma\ll1$ we get the value of the critical current $j_c$ for the vortex entry:
\begin{equation}
\label{sign}
\frac{j_c-j_c(\gamma=0)}{j_0}  \sim  \pm\gamma \ .  
\end{equation}
The sign of  the correction, proportional to $\gamma$, depends on the signs of the spin--orbit interaction constant $\varepsilon_3$ and of the mixed product $[\mathbf n\times\mathbf h]\cdot\mathbf j$.
The sign change in the Eq.(\ref{sign}) occurs when either the direction of the current or the direction of the exchange field change to the opposite.
Experimentally the value of the above diode effect in critical current can be of the order of tens of percent (see \cite{,Gaggioli-PRR-2024} and refs. therein), which makes the described effects observable in S/F systems with spin-orbit interaction.

\section{Anisotropy of the vortex dynamics\label{sec:motion}}

\begin{figure}[t!]
\includegraphics[width=0.8\linewidth]{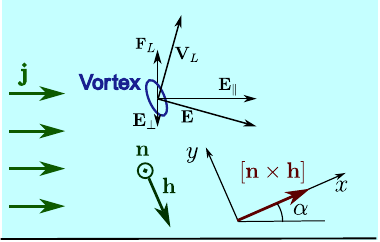}
\caption{Schematic picture illustrating the motion of a single vortex in a superconducting film in the presence of transport current  $\mathbf j$. $\mathbf F_L=c^{-1}\Phi_0[\mathbf j\times\mathbf z_0]$ is a Lorentz force, $\mathbf V_L$ is the vortex velocity and $\mathbf E$ is an induced electric field.}
\label{fig:hall}
\end{figure}

Our analysis of the vortex structure and vortex interactions in previous sections was essentially based on consideration of the regions far from the vortex cores.
Certainly, the odd-gradient terms in the above GL theory can result in the core distortions as well. We now proceed with a brief discussion of the effects arising from these distortions. 
As noted in Sec. \ref{sec:model}, the exclusion of the  linear SOI term leads to the effective mass anisotropy which should result in the distortion of the vortex core determined by the product $\varepsilon_1\varepsilon_3$.
The vortex core becomes elliptical leading to a number of well known consequences for experimentally measurable quantities \cite{Blatter-RMP-1994,Gorkov-UFN-1975,Kopnin-book-2001}.
In particular, the
vortex core anisotropy is known to affect the vortex dynamics in a resistive state resulting in the anisotropy of the flux-flow resistivity. 

The core anisotropy, described by the anisotropic mass tensor $\hat m$, can be derived directly from the Eq. \ref{Eq_GLeq1}.
In a coordinate system with $x$ axis along the $[\mathbf n\times\mathbf h]$ direction (Fig. \ref{fig:hall}) $\hat m$ is diagonal with components $m_x=m_0(1-3\mu), m_y=m_0(1-\mu)$, where $m_0$ is an standard GL effective mass (for $\varepsilon_3=0$) and $\mu=2\gamma\varepsilon_1 h/(\alpha\xi)\approx 2\gamma q\xi$.
From these expressions we can obtain the mass anisotropy $m_x/m_y=1-2\mu$, which results in the anisotropy of the vortex core profile. 
The current flowing in a superconductor produces a Lorentz force acting on a vortex which leads to the following equation for the vortex motion:
\begin{equation}
\hat\eta\mathbf V_L=\frac{\Phi_0}{c}[\mathbf j\times\mathbf z_0],
\end{equation}
where $\hat\eta$ is a viscosity tensor and $\mathbf V_L$ is the vortex velocity. The viscosity tensor can be derived from the time dependent GL theory following the approach described in \cite{Gorkov-UFN-1975} (see appropriate calculations in Ref.~\cite{GenkinJETP}).
The vortex motion, in turn, induces the electric field $\mathbf E=c^{-1}[\mathbf B\times\mathbf V_L]$, where $\mathbf B$ is the average magnetic field. For a magnetic field along the $z$ axis we get:
\begin{equation}
\mathbf E=\frac{\Phi_0B_z}{c^2}\left(
\frac{\mathbf x_0j_x}{\eta_y}+\frac{\mathbf y_0j_y}{\eta_x}
\right).
\end{equation}
In isotropic case the viscosity tensor is given by the relation $\eta_{ij}=\Phi_0^2\sigma_n/(2\pi c^2\xi^2)\delta_{ij}$, where $\sigma_n$ is a normal state conductivity and $\delta_{ij}$ is the Kronecker symbol. 
The anisotropy of the effective mass leads to the anisotropy of viscosity: for a small value $\mu\ll1$ we get
$\eta_x/\eta_y-1\sim\mu$ \cite{GenkinJETP,Hao-IEEE-1991}.

As a result, considering the vortex motion in a transport current carrying stripe (see Fig. \ref{fig:hall}) one can get the voltage in a transverse direction (perpendicular to the current) provided the exchange field is not directed exactly perpendicular or parallel to the current direction. 
The ratio of transverse and longitudinal electric field components is given by the expression:
\begin{equation}
    E_\perp /E_\parallel\sim \mu \sin\alpha \cos\alpha,
\end{equation}
where $\alpha$ is the angle between the exchange field and current direction.
The same value of anisotropy should reveal itself in the measurements of longitudinal and transverse voltages in  standard transport experiments.
Taking $q\xi\sim\gamma\sim 0.1$ we find a measurable value of anisotropy ratio $\mu\sim 0.01$.  
Note finally, that the mass tensor anisotropy can also arise from  the higher order terms in the $\gamma$ parameter, see Section \ref{sec:numerical} for details.

\section{Numerical simulation}\label{sec:numerical}

Below we solve the time-dependent Ginzburg-Landau equations with additional linear and cubic gradient terms in the dimensionless form.
We introduce the parameters $\eta=\varepsilon_{1}h/(\alpha\xi)$ and $\gamma=\varepsilon_{3}h/(\alpha\xi^{3})$ measuring the strengths of the linear and cubic terms, respectively.
The non-stationary GL theory here is used only as an approach which allows to get the numerical solution of the stationary equations corresponding to the free energy minimum (see also \cite{Sadovskyy-JCP-2015,Kato-PRB-1993,Kopasov-PRB-2023}).
The goal of these numerical simulations is to illustrate the reliability of the above approximate stationary analytical solutions which are valid only outside the vortex core and are based on the perturbative approach (small values of the third-order gradient terms are assumed). Besides, the numerical simulations allow us to investigate the order parameter behavior at any distances from the vortex center including the core region.
Three ingredients for numerical simulations are required: (i) the evolution equation for the order parameter, (ii) the expression for the transport current, and (iii) the boundary conditions that guarantee zero normal current at the sample edge. They are presented bellow. For convenience, we will write each of them separately, defining all the parameters, sometimes repeating in relation to what was written earlier. The evolution of the complex order parameter $\psi(\mathbf r,t)$ is governed by the equation
\begin{align}
u\!\left(\partial_t+i\mu\right)\psi &=
\bigl(1-|\psi|^{2}\bigr)\psi
+(\nabla-i\mathbf A)^{2}\psi             \nonumber\\
&\quad+2i\eta\,\mathbf x_{0}\!\cdot\!(\nabla-i\mathbf A)\psi          \nonumber\\
&\quad-2i\gamma\,(\nabla-i\mathbf A)^{2}
             \!\bigl[\mathbf x_{0}\!\cdot\!(\nabla-i\mathbf A)\bigr]\psi,
\label{eq:tdgl}
\end{align}
where $u$ is the dimensionless relaxation constant, $\mu$ is the electrochemical potential (which equals to zero in equilibrium), $\mathbf A=[\mathbf B\times\mathbf r]/2$ is the vector potential of a uniform perpendicular field $\mathbf B$, $\mathbf x_{0}=[\mathbf n\times\mathbf h]=(1,0)$ is fixed by the exchange field. We emphasize that here we assume the transverse dimensions of the sample to be much less than the effective screening length $\lambda_{eff}$ and, thus, we neglect the magnetic field inhomogeneity caused by the screening effects.

The gauge-invariant supercurrent entering Maxwell’s equations is  
\begin{align}
\mathbf J &=
\mathrm{Im}\!\Bigl[\psi^{*}\bigl(\nabla-i(\mathbf A-\eta\mathbf x_{0})\bigr)\psi\Bigr]  \nonumber\\
&\quad+\gamma\,\mathrm{Re}\!\Bigl[
(\nabla+i\mathbf A)\psi^{*}\,[\mathbf x_{0}\!\cdot\!(\nabla-i\mathbf A)]\psi   \nonumber\\
&\qquad\qquad\;
-\psi^{*}(\nabla-i\mathbf A)[\mathbf x_{0}\!\cdot\!(\nabla-i\mathbf A)]\psi     \nonumber\\
&\qquad\qquad\;
-\psi^{*}\mathbf x_{0}(\nabla-i\mathbf A)^{2}\psi\Bigr].
\label{eq:current}
\end{align}

The conditions at the sample boundary $S$ which are consistent with the absence of current through the boundary directly follow from varying the GL functional:
\begin{align}
&\mathbf N\!\cdot\!i\gamma(\nabla-i\mathbf A)\psi\bigl|_{S}=0,
\label{eq:bc1}\\
&\mathbf N\!\cdot\!\Bigl[(\nabla-i(\mathbf A-\eta\mathbf x_{0}))
+\,i\gamma(\nabla-i\mathbf A)[\mathbf x_{0}\!\cdot\!(\nabla-i\mathbf A)]
\nonumber\\
&\qquad+\,i\gamma\mathbf x_{0}(\nabla-i\mathbf A)^{2}\Bigr]\psi\bigl|_{S}=0.
\label{eq:bc2}
\end{align}

\begin{figure}
\begin{minipage}[H]{1\linewidth}
\center{\includegraphics[width=1\linewidth]{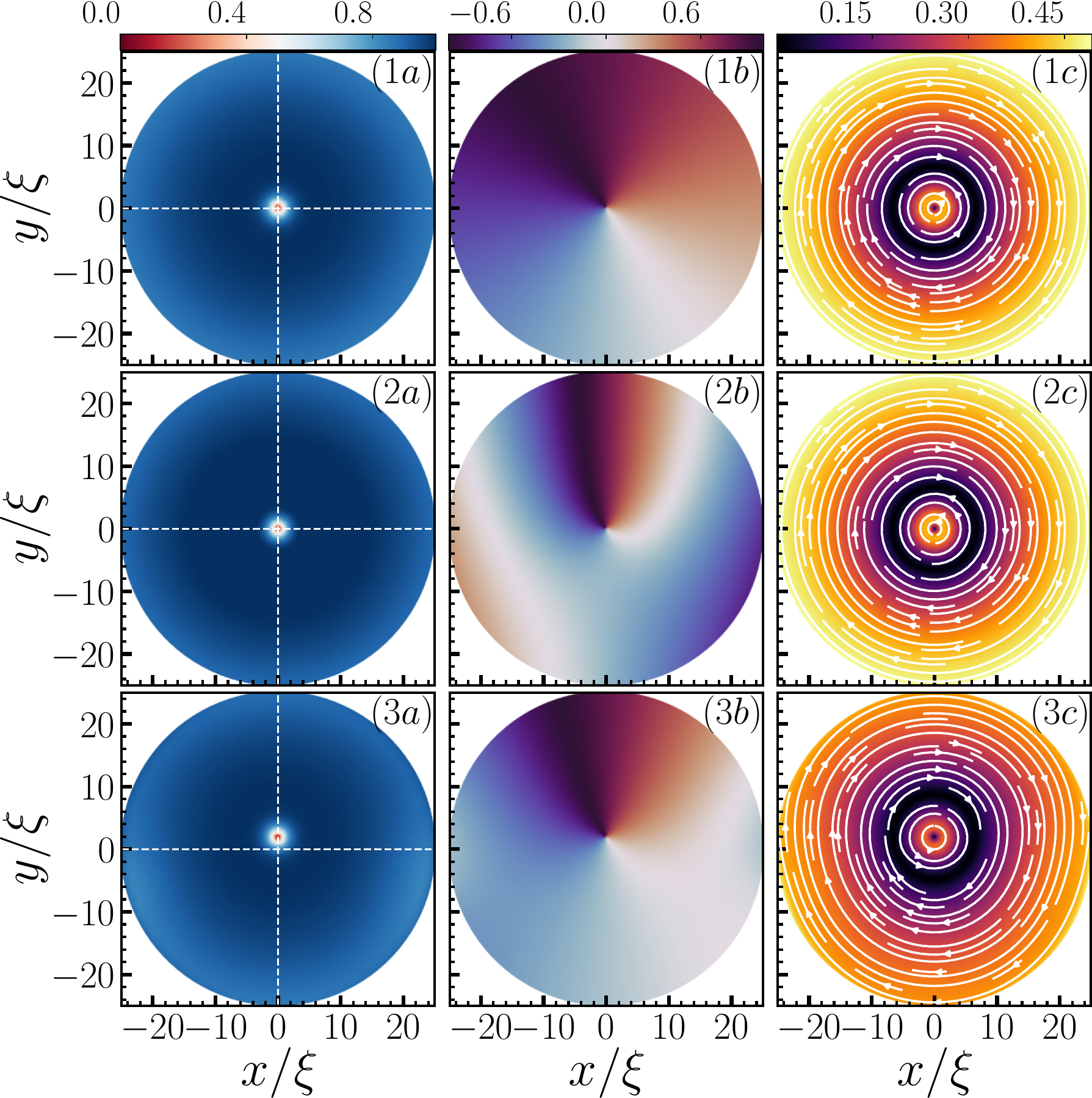}}
\end{minipage}
\caption{Numerical vortex profiles in a disc of radius $R = 25\xi$ at the reduced field $B = 0.12\,B_{c2}$.  
Rows: (1) conventional GL; (2) GL with a linear spin-orbit term, GL$(0.25,0)$; (3) GL with a cubic term, GL$(0,0.025)$.  
Columns: (a) modulus $|\psi|$, (b) phase $\theta/\pi$, (c) current-density magnitude $|\mathbf J|$ (white streamlines trace the current direction).}
\label{Three cases}
\end{figure}

We do not study here the details of relaxation processes and omit the potential $\mu$ in the above equations keeping in mind that
the solution of the dynamic equations is needed only to find the final stationary state with  minimal energy. Thus, the equations~\eqref{eq:tdgl}-\eqref{eq:bc2} are integrated in real time until a stationary state ($\partial_t\psi\!\approx\!0$) is reached.

The lengths are expressed in coherence-length units~$\xi$, magnetic fields
in $B_{c2}=\Phi_{0}/(2\pi\xi^{2})$, and current density in $4\xi B_{c2}/(10\mu_{0}\lambda^{2})$. For simplicity, we denote the equilibrium solution corresponding to $\eta = 0, \gamma = 0$ as GL solution and the result for $\eta \neq 0, \gamma \neq 0$ as GL($\eta, \gamma$) one.

The open-source \textsc{pyTDGL} package provides a baseline
second-order-accurate solver~\cite{pytdgl}. The existing accuracy of the procedure ($\mathcal{O}(h^2)$) is not enough to solve the modified TDGL equation due to the term with the third derivative. For stable solution of~\eqref{eq:tdgl}-\eqref{eq:bc2} in a wide range of $\gamma$, we implemented least square-based finite difference scheme (LSFD) for calculating space derivatives with $\mathcal{O}(h^4)$ accuracy \cite{lsq}, where $h$ is a characteristic edge's length of the triangular mesh. The description of the LSFD method is presented in Appendix~\ref{sec:app_b} and~\ref{sec:app_c}. Further, we proceed to discuss the results of specific numerical simulation.

\subsection{Single Vortex for GL($\eta, 0$)}

We begin by testing whether the \textit{linear} spin-orbit contribution, set by the parameter $\eta$, alters an isolated vortex. With $\gamma=0$ Eqs.~(\ref{eq:tdgl})-(\ref{eq:bc2}) simplify after the gauge shift $\mathbf{\bar A}\equiv\mathbf A-\eta\mathbf x_{0}$, giving the standard
time-dependent GL system
\begin{subequations}
\label{motion}
\begin{align}
&\!u\partial_t \psi\! =\! \left( 1 + (\eta \textbf{x}_{0})^2\! -\!  |\psi|^2 \right) \psi +  (\nabla - i\bar{\textbf{A}})^2 \psi = 0,\label{eqn:psi_a_1}\\
&\textbf{N} \cdot (\nabla - i\bar{\textbf{A}} )\psi|_{S} = 0,\label{eqn:psi_a_2}\\
&\textbf{J}  =  \text{Im} \, \Big(  \psi^{*}(\nabla - i\bar{\textbf{A}}) \psi \Big).\label{eqn:psi_a_2}
\end{align}
\end{subequations}
Because the transformation is purely gauge, it can only add a uniform phase to~$\psi$; the vortex position and current pattern must remain unchanged.

\begin{figure}
\begin{minipage}[H]{1\linewidth}
\center{\includegraphics[width=1\linewidth]{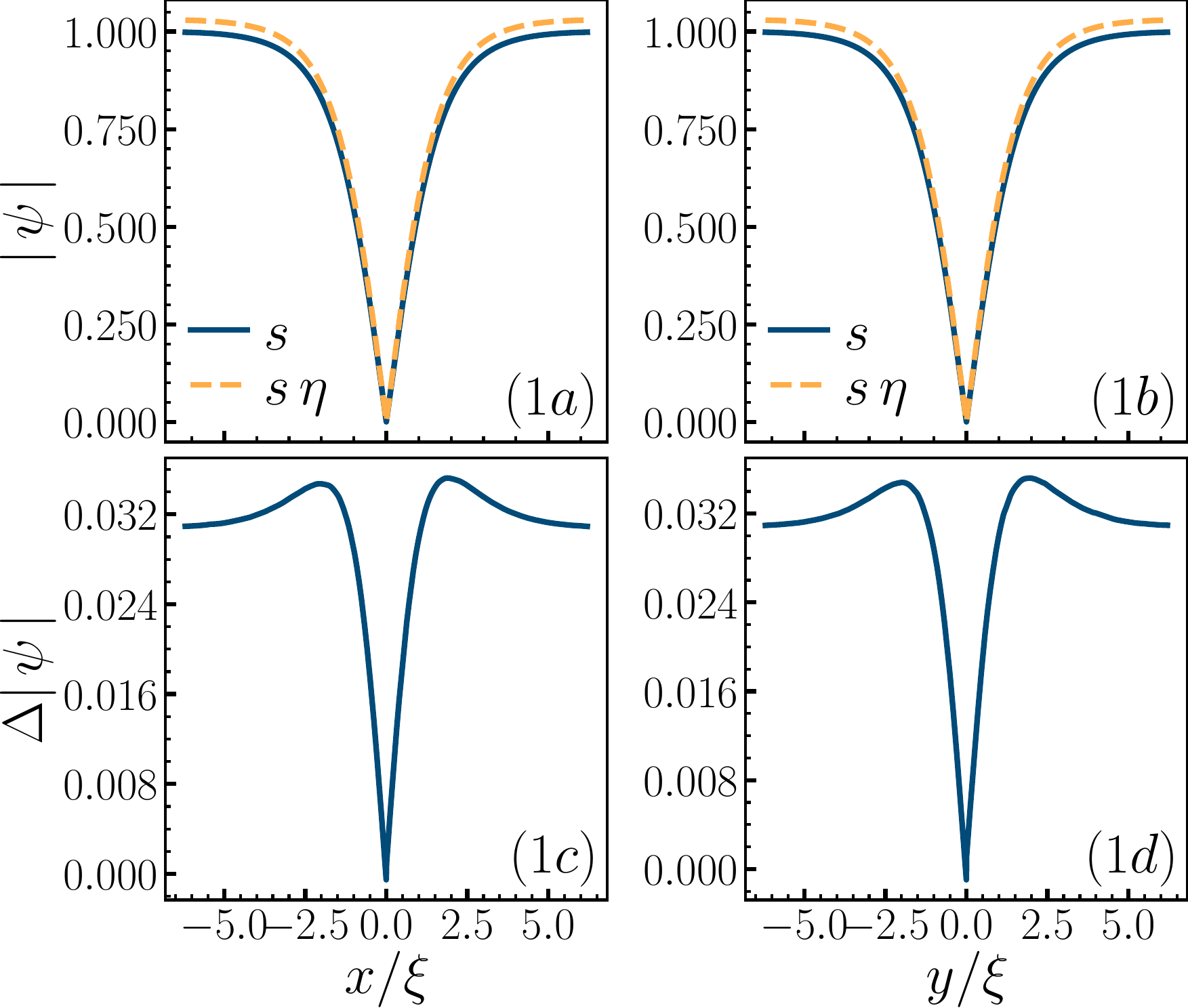}}
\end{minipage}
\caption{Comparison of the order-parameter amplitude in the vicinity of the vortex core. The core center is taken as the coordinate origin.  
All curves are obtained from time-dependent GL simulations for a single vortex in a disc of radius $R = 25\xi$ at the reduced field $B = 0.12\,B_{c2}$. Panel (1a) show the dependence of the order parameter's magnitude on the $x$ coordinate. Panel (1b) shows the dependence of the order parameter's magnitude on coordinate $y$. Numerical simulation for GL case is marked as \textit{s}. Numerical simulation for GL(0.25, 0) case is marked by  $s\;\eta$. Panels~(1c) and~(1d) plot the differences $\Delta|\psi|$ along $x$ and $y$, respectively, emphasising the small correction produced by the linear gradient term.}
\label{Gl lin psi}
\end{figure}

\begin{figure}
\begin{minipage}[t!]{1\linewidth}
\center{\includegraphics[width=1\linewidth]{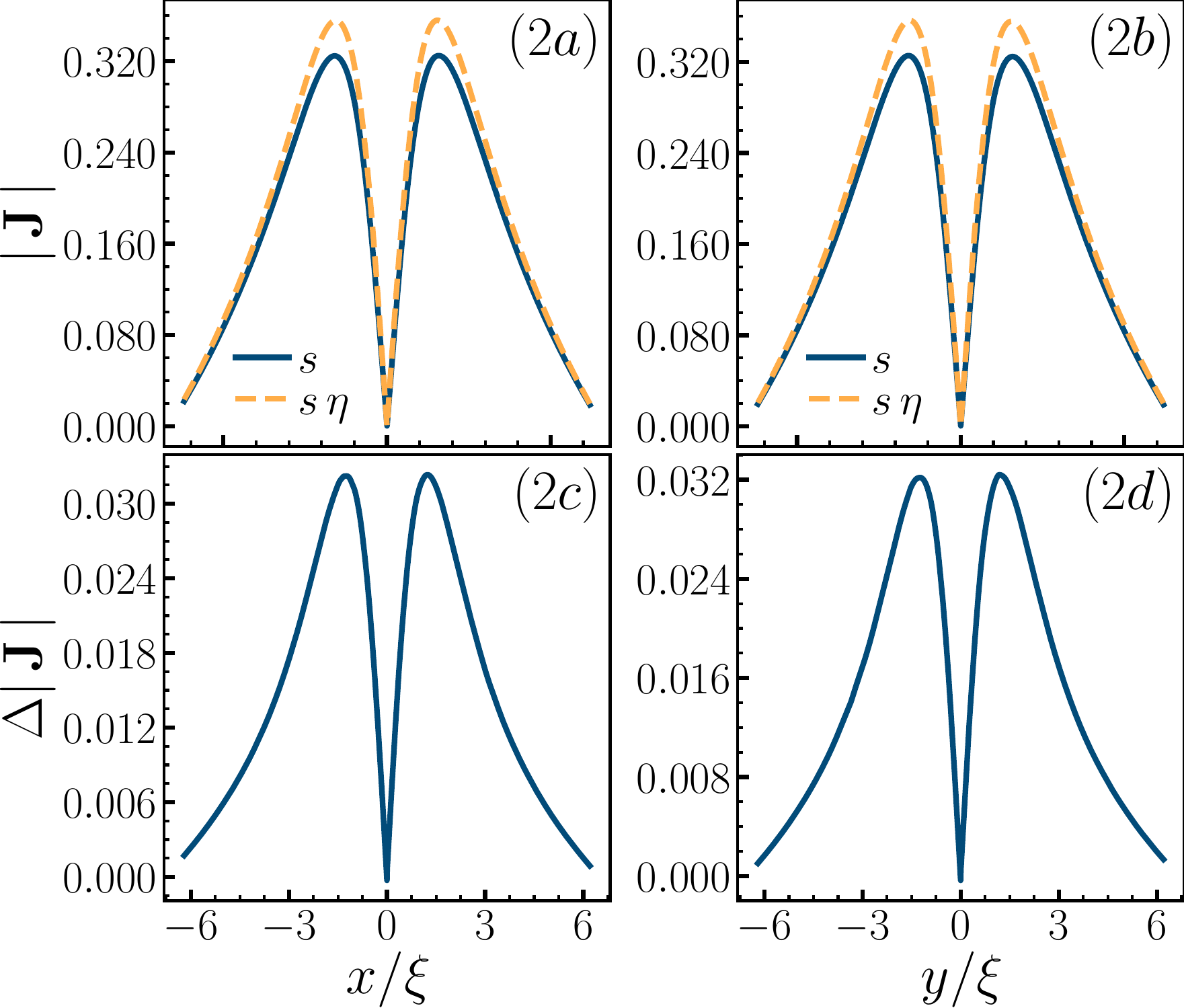}}
\end{minipage}
\caption{Comparison of the super-current density in the vicinity of the vortex core. The core center is chosen as the origin of coordinates.  All curves are obtained from time-dependent GL simulations for a single vortex in a disc of radius $R = 25\xi$ at the reduced field $B = 0.12\,B_{c2}$.  Panel (2a) shows the profile of $|\mathbf J|$ along the $x$ axis, while panel (2b) shows the profile along the $y$ axis.  Results for the reference GL model are labeled by the letter \textit{s}; results for $\text{GL}(0.25,0)$ are labelled as $s\,\eta$.  Panels (2c) and (2d) plot the differences $\Delta|\mathbf J|$ along $x$ and $y$ directions, respectively, emphasising the correction generated by the linear-gradient term. }
\label{Gl lin J}
\end{figure}

Numerical results confirm this conclusion. Figures~\ref{Three cases}(1) and~\ref{Three cases}(2) show that the vortex obtained for GL($0.25,0$) is identical to the reference GL solution,
apart from the overall phase twist. The linear term simply rescales the condensate density: both $|\psi|$ and $|\mathbf J|$ are enhanced by $\sim\!\eta^{2}/2$ (Figs.~\ref{Gl lin psi}, \ref{Gl lin J}),
effectively reducing the coherence length. This homogeneous renormalisation can be removed by normalising the order
parameter and does not produce any chiral distortion of the vortex.

We therefore conclude that the linear-gradient invariant does not introduce any qualitative changes to either the vortex core position or the circular current flow.  All non-trivial chiral effects discussed
below originate from the cubic gradient term~$\gamma$.

\subsection{Single Vortex for GL($0, \gamma$)}
\begin{figure}
\begin{minipage}[H]{1\linewidth}
\center{\includegraphics[width=1\linewidth]{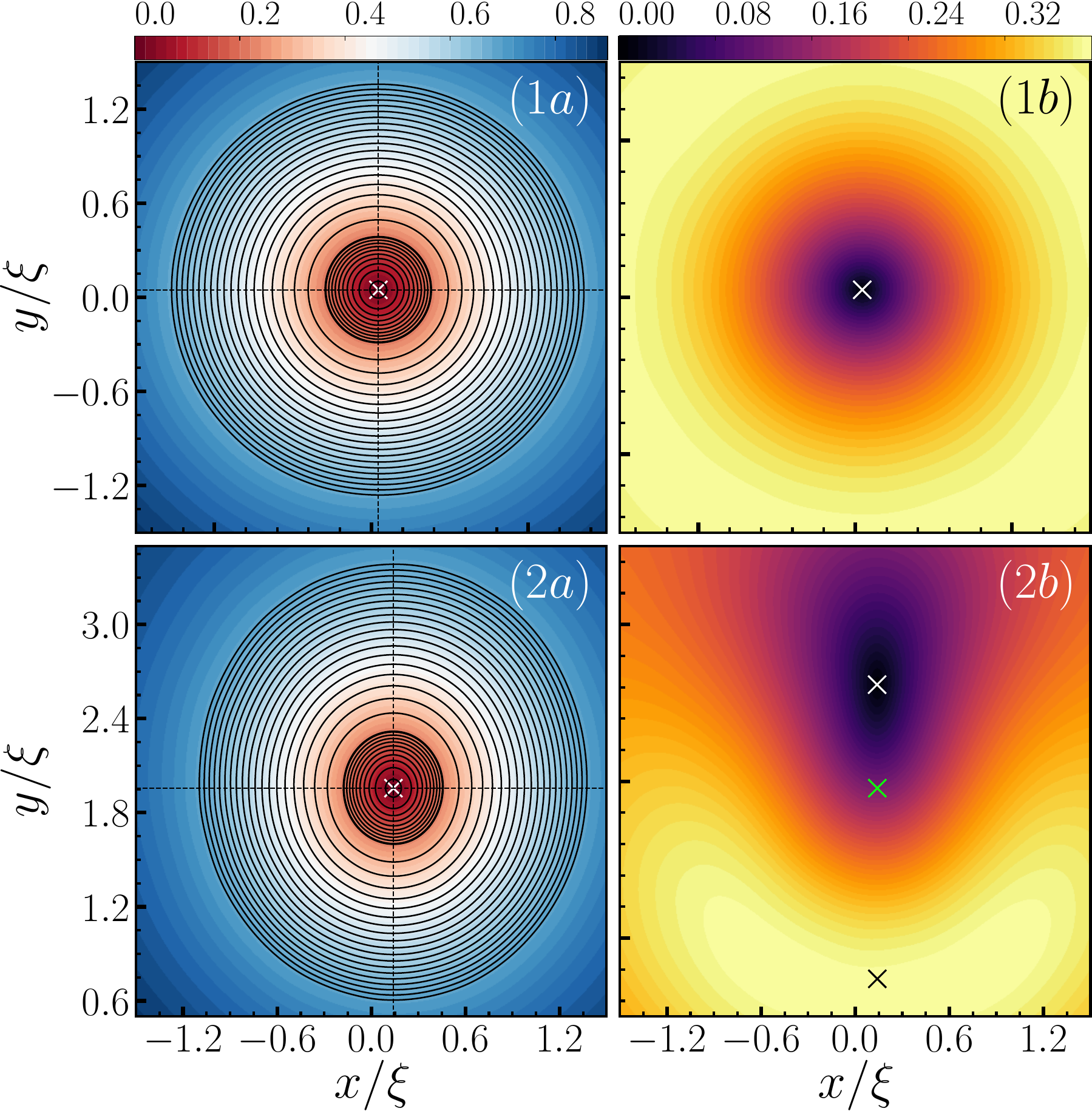}}
\end{minipage}
\caption{%
High-resolution maps of the order-parameter amplitude (1a), (2a) and the supercurrent‐density modulus (1b), (2b) for a single vortex, obtained from time-dependent GL simulations in a circular film of radius $R = 25\xi$ at the reduced field $B = 0.06\,B_{c2}$. Panels (1a) and (1b) show the reference GL solution, whereas (2a) and (2b) correspond to the GL$(0,0.3)$ model with cubic spin–orbit coupling $\gamma = 0.3$.  
Cross symbols highlight key points: in each $|\psi|$ map the white cross marks the minimum amplitude; in the GL$(0,0.3)$ current map the white cross marks the minimum current density, the green cross the minimum of $|\psi|$, and the black cross the current maximum.  
The cubic term shifts the vortex core by $\simeq 2\xi$ along $+\hat y$, produces distortion of the core shape into an elliptical form, and introduces a pronounced up–down asymmetry in the current profile.%
}
\label{Vortex_pic}
\end{figure}
\begin{figure}
\begin{minipage}[H]{1\linewidth}
\center{\includegraphics[width=1\linewidth]{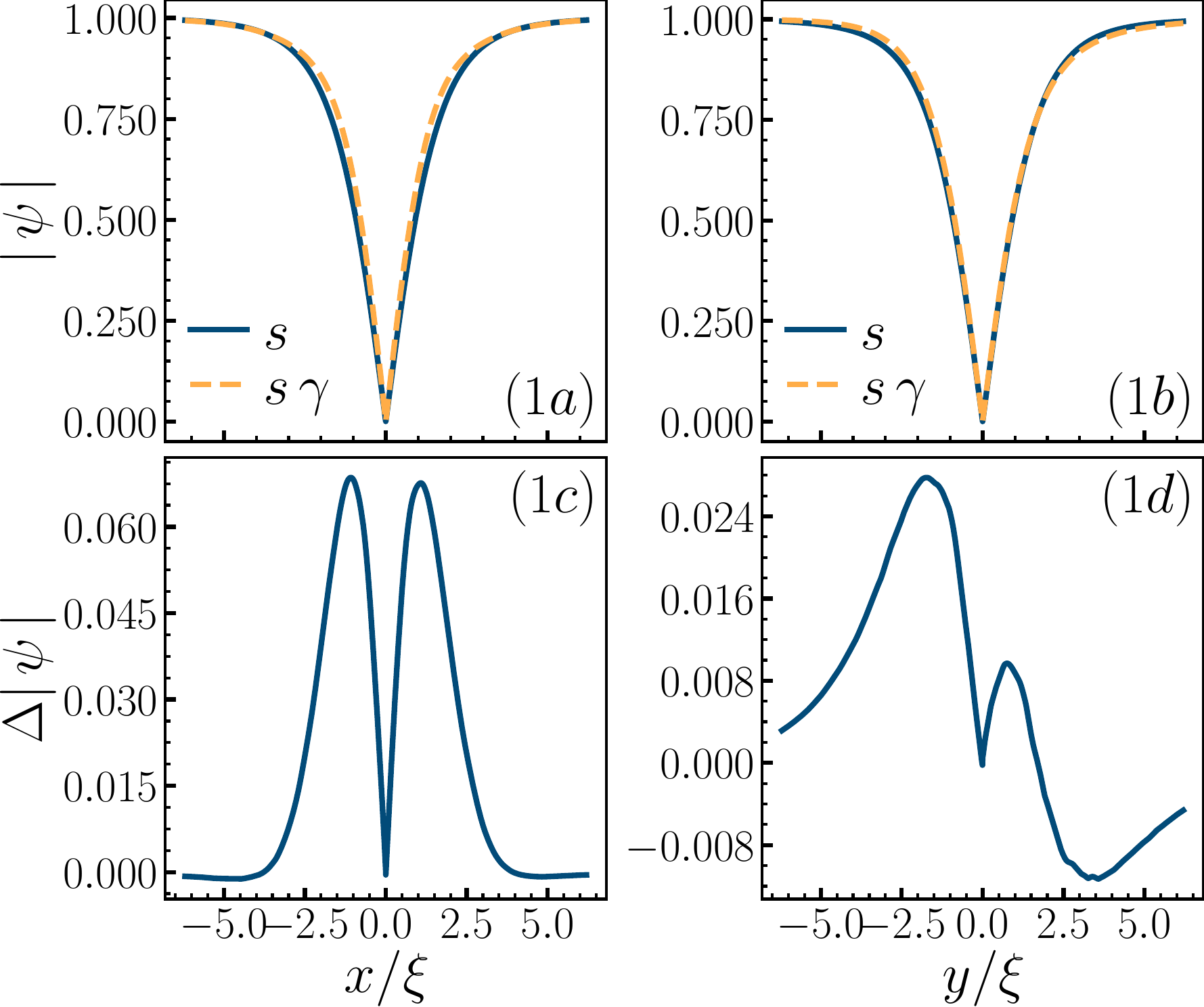}}
\end{minipage}
\caption{%
Line cuts of the order-parameter amplitude $|\psi(x,y)|$ taken through the vortex centre for the same simulation set as in Fig.~\ref{Vortex_pic}.  
A single vortex is relaxed in a disc of radius $R = 25\xi$ at the reduced field $B = 0.06\,B_{c2}$. The core center is chosen as the origin of coordinates. 
Panels (1a) and (1b) display $|\psi|$ along the $x$ and $y$ axes, respectively: solid line \textit{s} correspond to the reference GL solution, dashed line \textit{s}\,$\gamma$ to the GL$(0,0.3)$ model with cubic gradient terms. Panels (1c) and (1d) plot the point-wise difference $\Delta|\psi|$ along the same cuts, highlighting the distinct modulation introduced by the cubic term.  Apart from this small correction the radial profiles remain virtually identical, indicating that the $\simeq 2\xi$ vortex shift observed in Fig.~\ref{Vortex_pic} occurs without appreciable broadening or contraction of the vortex core.}
\label{GL_03_psi}
\end{figure}
\begin{figure}
\begin{minipage}[H]{1\linewidth}
\center{\includegraphics[width=1\linewidth]{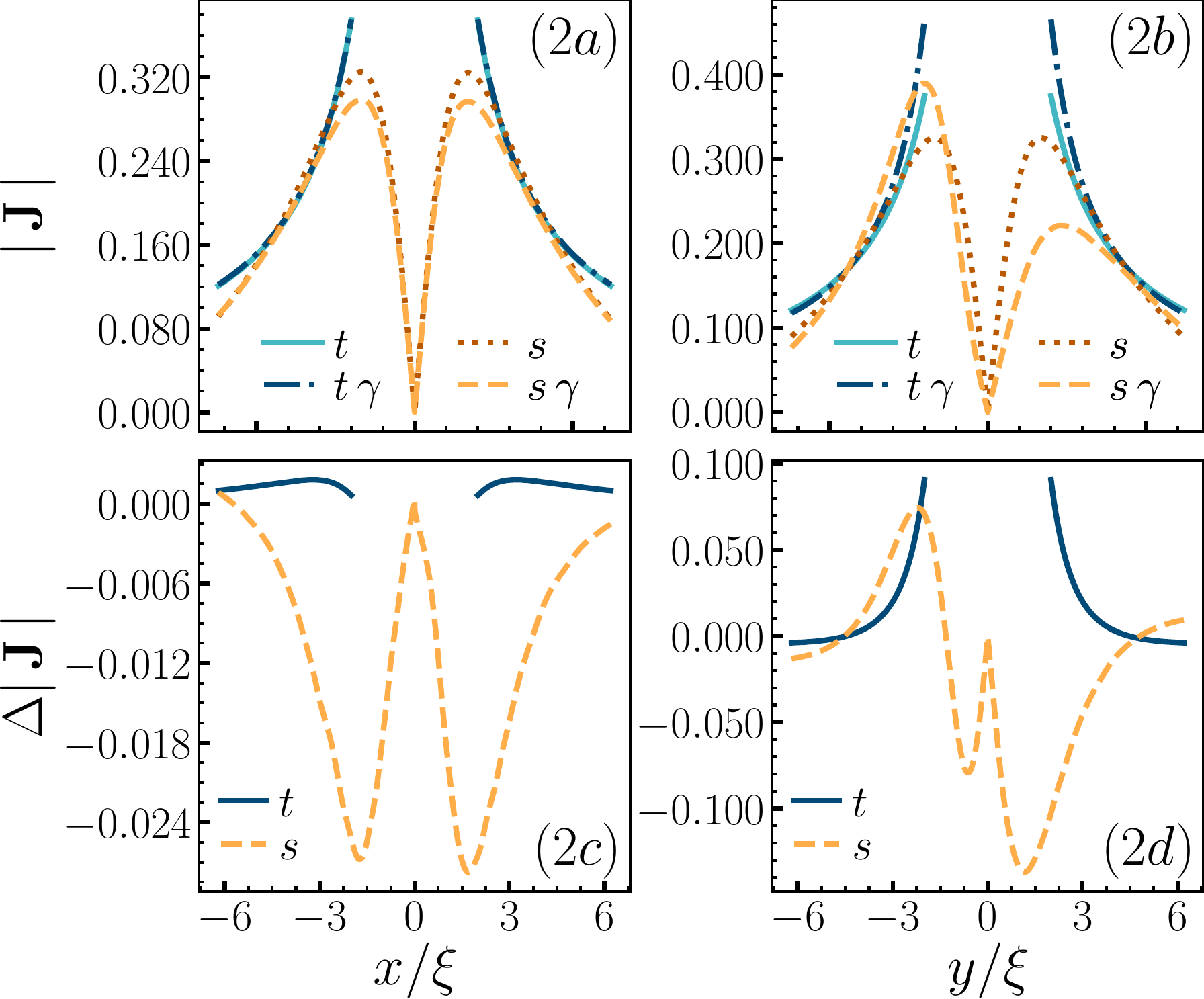}}
\end{minipage}
\caption{Line profiles of the supercurrent modulus $|{\bf J}(x,y)|$ extracted from the same simulations as in Fig.~\ref{Vortex_pic} ($R = 25\xi$, $B = 0.06\,B_{c2}$). Result was obtained with numerical simulation of a single vortex in the disk with $B=0.06 B_{c2}$ and $R=25 \xi$. The core center is chosen as the origin of coordinates. Panels (2a) and (2b) plot $|{\bf J}|$ along the $x$ and $y$ axes, respectively. Estimates of perturbation theory for GL case is marked as $t$. Numerical simulation for GL case is marked as $s$. Estimates of perturbation theory for GL(0, 0.3) case is marked as $t\,\gamma$. Numerical simulation for GL(0, 0.3) case is marked as $s\,\gamma$. Panels (2c) and (2d) display the point-wise differences $\Delta|{\bf J}|$ along $x$ and $y$, comparing theory $t$ with theory $t\, \gamma$ and simulation $s$ with simulation $s\,\gamma$; the narrow spike at the origin reflects the logarithmic divergence of the analytic current near the vortex core.  
Apart from this core singularity, the numerical and analytical curves coincide, confirming that the cubic term mainly redistributes the angular current pattern without affecting the overall radial decay.}
\label{GL_03_J}
\end{figure}

To isolate the effect of the cubic gradient term we set $\eta=0$ and retain only $\gamma\neq0$ in the modified TDGL theory. The resulting evolution equation, boundary conditions, and current expression read
\begin{subequations}
\label{motion_B}
\begin{align}
&\!\!\!\!u\partial_{t}\psi =
(1-|\psi|^{2})\psi
+(\nabla-i\mathbf A)^{2}\psi\nonumber \\
&\qquad\qquad
-2i\gamma(\nabla-i\mathbf A)^{2}
          \bigl[\mathbf x_{0}\!\cdot\!(\nabla-i\mathbf A)\bigr]\psi=0 ,\label{cubs1}\\
&\!\!\!\!\mathbf N\!\cdot\!(\nabla\!-\!i\mathbf A)\psi\bigl|_{S}=0, \label{cubs2}\\
&\!\!\!\!\mathbf N\!\cdot\!
\bigl\{(\nabla\!-\!i\mathbf A)
      \left[\mathbf x_{0}\!\cdot\!(\nabla\!-\!i\mathbf A)\right]
      \!+\!\mathbf x_{0}(\nabla\!-\!i\mathbf A)^{2}\bigr\}
      \psi\bigl|_{S}=0 ,
\label{cubs2_1}\\
&\!\!\!\!\mathbf J\! =\! \operatorname{Im}
   \bigl[\psi^{*}(\nabla\!-\!i\mathbf A)\psi\bigr]
\!+\!\gamma\operatorname{Re}
   \bigl[(\nabla\!+\!i\mathbf A)\psi^{*}(\mathbf x_{0}\!\cdot\!(\nabla-i\mathbf A))\psi\nonumber\\
&\!\!\!-\psi^{*}(\nabla\!-\!i\mathbf A)
         (\mathbf x_{0}\!\cdot\!(\nabla\!-\!i\mathbf A))\psi
        \!-\!\psi^{*}\mathbf x_{0}(\nabla\!-\!i\mathbf A)^{2}\psi\bigr].\!
\label{cubs3}
\end{align}
\end{subequations}
Comparison of the baseline GL vortex with its cubic-gradient counterpart GL$(0,0.025)$ is summarized in Fig.~\ref{Three cases}, panels (1) and (3). The two data sets were obtained under identical external conditions ($B=0.12\,B_{c2}$, $R=25\xi$), so any difference can be ascribed solely to the presence of the cubic term. The main difference between the GL and GL$(0,0.025)$ vortices is a minor vortex displacement on Fig.~\ref{Three cases} (3a). For a more comprehensive review of the vortex features, we increase the parameter $\gamma$ and consider the case GL$(0,0.3)$. A first inspection of the color maps in Fig.~\ref{Vortex_pic}(1a) and \ref{Vortex_pic}(2a) reveals that the amplitude profile $|\psi(\mathbf r)|$ is distorted and the vortex core has an elliptical form. With increasing $\gamma$ the magnitude of displacement likewise increases. This shift coincides, within the numerical resolution, with the analytical prediction following from Eq.~(\ref{shift teor}). For detailed analysis of vortex distortion we plot cross-sections through the vortex center in Fig.~\ref{GL_03_psi}(1a)-(1d). The solid and dashed curves representing GL and GL$(0,0.3)$ overlap almost perfectly, while the residual differences, shown separately in panels (1c) and (1d), appeared to be quite different. Panel (1c) shows symmetrical distinction along x-axis, causing an elliptical distortion of the vortex core. Along the y-axis on panel (1d) the distinction is asymmetric and notably weaker, yet it extends far beyond the core, forming a long-range background signature.

The impact on the supercurrent is far more pronounced. Figures \ref{Vortex_pic}(1b) and \ref{Vortex_pic}(2b) demonstrate that the azimuthal flow pattern characteristic of an isolated GL vortex becomes strongly skewed once $\gamma$ is switched on. Under the core (along the $y$ axis) the streamlines crowd together, producing an enhanced local current density, whereas on the opposite side they fan out. This asymmetry is exactly what one expects from the chiral correction in Eq.~\eqref{cubs3}, which adds a term proportional to $\mathbf x_{0}\cdot\mathbf v$ and therefore amplifies (suppresses) the current where $\mathbf v$ is parallel (antiparallel) to $\mathbf x_{0}$. A detailed comparison of azimuthal profiles is given in Fig.~\ref{GL_03_J}(2a)-(2d); outside the immediate core region the numerical data coincide with the perturbative formula~(\ref{j_teor}) within the plotting accuracy.

An additional and subtler consequence of chirality is the decoupling of the minima of $|\psi|$ and $|\mathbf J|$. For a conventional GL vortex these two points coincide at the core centre, but in the diode-type model the current minimum is shifted \emph{upwards} by approximately $\gamma\xi$, as marked by the white and gold crosses in Fig.~\ref{Vortex_pic}(2b).  

This separation follows from  in Eq.~\eqref{cubs3}; maximizing its magnitude drives the current minimum away from the core one.   The shift is clearly reproduced in the difference plots of Fig.~\ref{GL_03_J}, panels (2c) and (2d). Physically, the appearance of two distinct centers, i.e., a phase-winding center and a minimum current center, implies that a vortex subject to the cubic gradient interaction carries an intrinsic dipole moment. For the single-vortex problem treated here by numerical simulation, the cubic term thus produces several robust signatures: a rigid core shift linear in~$\gamma$, a chiral deformation of the current loops scaling as $\gamma$, a relative displacement of the current and amplitude minima also scaling as~$\gamma$, and elliptical distortion of vortex core.

\subsection{Magnetic field dependence of the vortex shift }

\begin{figure}[t]
\includegraphics[width=\linewidth]{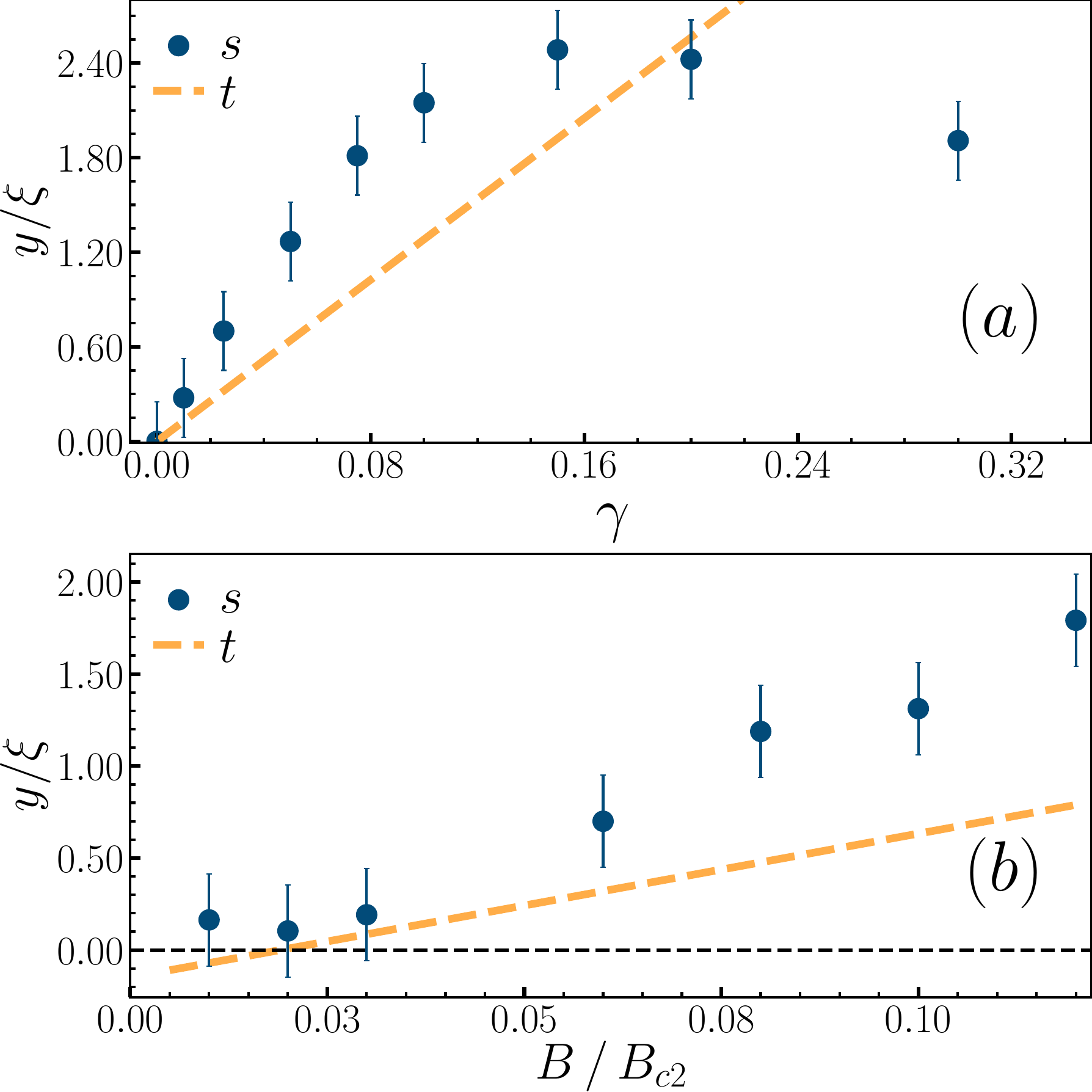}
\caption{
(a) Vortex displacement $\Delta y$ versus cubic coefficient $\gamma$ at $B=0.06\,B_{c2}$ for a disc of radius $R=12.5\xi$. (b) $\Delta y$ versus perpendicular field for fixed $\gamma=0.025$ and $R=25\xi$.  
Label (\textit{s}) corresponds to numerical data; dashed lines (\textit{t}) are perturbative predictions.
Error bars correspond to $0.5\xi$, the mesh spacing in the simulations.}
\label{pic shift}
\end{figure}

A key goal of our numerical study is to establish the $\gamma$–range in which the perturbative prediction for the vortex displacement remains quantitative. Figure~\ref{pic shift}(a) plots the core shift $\Delta y$ versus $\gamma$ at fixed field $B=0.6\,B_{c2}$. This field value is chosen because the cubic term modifies the Bean–Livingston barrier [Eq.~\eqref{bin correction}] and hence the lower critical field $B_{c1}$.  
For a disc of radius $R=25\xi$ we obtain $B_{c1}\simeq0.15\,B_{c2}$ in the pure GL model, while the cubic gradient lowers that threshold; selecting $B=0.6\,B_{c2}$ therefore guarantees a single, stable vortex over the full $\gamma$ interval explored.

Perturbation theory predicts that the shift depends linearly on both $\gamma$ and the perpendicular field,
\begin{equation}
\Delta y=\gamma\Bigl[-2\ln\!\frac{R}{\xi}+0.5+B\frac{R^{2}}{2\xi^{2}}\Bigr],
\label{shift teor dim}
\end{equation}
the dimensionless form of Eq.~\eqref{shift teor}.  The straight‐line behaviour visible in Fig.~\ref{pic shift}(a) confirms this relation up to $\gamma\approx0.1$, beyond which higher‐order terms start to matter.
Figure~\ref{pic shift}(b) presents the complementary scan: $\Delta y$ as a function of $B$ at fixed $\gamma=0.025$. The numerical and analytical curves coincide for $B\gtrsim0.2\,B_{c2}$, validating the linear field dependence in this regime. At lower fields the perturbative curve changes sign, whereas the simulation shows a continued growth of the displacement -- an indication that, close to the Bean-Livingston threshold, the cubic term alters the surface barrier beyond the simple estimate of Eq.~\eqref{bin correction}.  
A refined treatment of this low‐field crossover lies outside the present scope and will be addressed elsewhere.

\section{Conclusion}

We have developed a comprehensive Ginzburg-Landau description of superconducting thin-film hybrid structures with intrinsic diode effect accounting for
two essential ingredients: an in-plane exchange field and an interfacial spin-orbit interaction.  A gauge transformation removes the linear gradient term and leaves a single dimensionless parameter $\gamma$ that multiplies a cubic spatial derivative; this compact one-parameter model captures all diode-related phenomena in the {\it Abrikosov} vortex state.

Starting from the generalized GL functional we derived closed-form expressions for (i) the chiral distortion of the superfluid velocity and current around an {\it isolated} vortex, (ii) the $\gamma$-induced non-central force and torque acting in a vortex-antivortex pair, and (iii) the direction-dependent correction to the Bean-Livingston entry barrier in mesoscopic geometries.  In particular, the cubic term was shown to shift vortex cores by an amount proportional to $\gamma$ and to generate a transverse force that can rotate an entire vortex ensemble showing clear signatures of broken inversion and time-reversal symmetry.  All analytical predictions were corroborated by time-dependent Ginzburg-Landau simulations implemented with a fourth-order least-squares finite-difference solver, yielding quantitative agreement over a moderate range of parameters.

These results supply the key fingerprints of vortex physics in superconductors with an intrinsic diode effect and outline concrete routes for experimental detection.  The predicted core shift and current asymmetry can be probed with scanning SQUID and tunneling microscopy, whereas the non-central intervortex forces and the modified Bean-Livingston barrier should manifest in anisotropic flux-entry fields and in the orientation of vortex configurations observed by Lorentz or magneto-optical imaging.  Beyond fundamental interest, the ability to manage vortices in systems with exchange interaction and spin-orbit coupling opens avenues toward non-reciprocal superconducting circuitry, and fluxonic logic.

Future work may address the collective dynamics of vortex lattices in the strong-screening limit, include microscopic corrections away from the Ginzburg-Landau domain, and explore how engineered spin-orbit and exchange field textures could be used to craft designer vortex potentials for superconducting electronics and topological qubits.

\begin{acknowledgments}

This work was supported by the Russian Science Foundation (Grant No. 25-12-00042) in part of the analytical calculations in Secs. III, IV A, IV B
and by the Grant of the Ministry of science and Higher education of the Russian Federation No. 075-15-2025-010 (in part of analysis of the Bean-Livingston barrier in Sec. IV C)
and by MIPT Project No. FSMG-2023-0014 (in part of the analysis of the anisotropic vortex dynamics in Sec. V).
The numerical simulations of A.K. (Section VI) have been supported by the Icelandic Research Fund (Ranns\'oknasj\'o{\dh}ur, Grant No.~2410550).
The work of A.I.B. was supported by ANR SUPERFAST and the
LIGHT S\&T Graduate Program.

\end{acknowledgments}


\appendix

\section{Calculation of the vortex energy in the disk}
For calculation of the energy (\ref{Eq_FE}) for a vortex in the disk we consider the expression 
\begin{multline}
v^2v_x=\frac{\sin\theta}{r^3}\left(1-\frac{br^2}{R^2}\right)^3-\frac{y}{r^2R^2}(1+2\sin^2\theta)\\
\times(1-\frac{br^2}{R^2})^2(1-b)+O(y^2).
\end{multline}
To calculate the total free energy, it is necessary to integrate its density over the superconducting disk. We  integrate over the circle shown by the dashed line in Fig.~\ref{fig integr}, add the integral over the crescent-shaped region I and subtract the integral over the crescent-shaped region II:
\begin{figure}[t]
\includegraphics[width=0.7\linewidth]{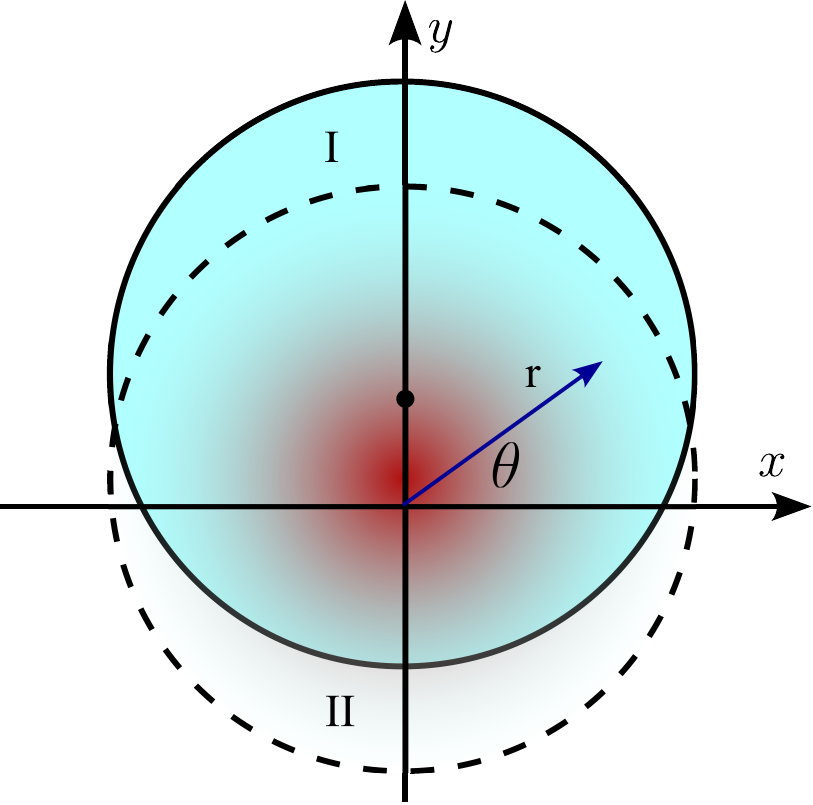}
\caption{
A schematic view of a superconducting disk in perpendicular magnetic field. The dashed circle and crescent-shaped regions I and II are used in calculation of the free energy.}
\label{fig integr}
\end{figure}
\begin{multline}
\int\limits_{disk}FdS\approx\int_0^{2\pi}\int_0^{R+y\sin\theta}F\,rdrd\theta
\approx\\
\approx\int\limits_0^{2\pi}\int\limits_0^{R}F\,rdrd\theta+yR\int\limits_0^{2\pi}F(R,\theta)\sin\theta d\theta.
\end{multline}
For symmetry reasons, only the term $\propto y$ will contribute to the first integral:
\begin{multline}
\int_0^{2\pi}\int_0^{R}v^2v_x\,rdrd\theta\\
=\frac{4\pi y(b-1)}{R^2}\int_0^{R}\frac{(R^2-br^2)^2}{R^4r}\,dr\\
=-\frac{4\pi y(1-b)}{R^2}\left(\ln\frac{R}{\xi}-b+b^2/4\right).
\end{multline}
The second integral is contributed by the main term independent on $y$, however the sickles area is proportional to $y$:
\begin{multline}
yR\int_0^{2\pi}v^2v_x\sin\theta d\theta\\=y\int_0^{2\pi}\frac{\sin^2\theta}{R^2}(1-b)^3 d\theta=\frac{\pi y}{R^2}(1-b)^3.
\end{multline}
Summing up two integrals, we obtain:
\begin{equation}
\int\limits_{disk}v^2v_x\,dS=
\frac{\pi y(1-b)}{R^2}\left(-4\ln R/\xi+1+2b\right).
\end{equation}
Now we can calculate the total energy:
\begin{multline}
U=\frac{dH_c^2\xi^2}{2}\left[\ln\frac{R}{\xi}+\ln(1-\frac{y^2}{R^2})-b(1-\frac{y^2}{R^2})\right.-\\
-\left.\frac{y\gamma\xi(1-b)}{R^2}\left(-4\ln\frac{R}{\xi}+1+2b\right)\right].
\end{multline}
Minimization of this energy with respect to parameter $y$ in the limit $y\ll R$ allows us to find its equilibrium value
\begin{equation}
y_0= \gamma\xi\bigl[-2\ln(R/\xi)+0.5+b\bigr].
\label{shift teor}
\end{equation}

\section{Mesh and dimensionless units}\label{sec:app_b}

\begin{figure}
\begin{minipage}[H]{1\linewidth}
\center{\includegraphics[width=1\linewidth]{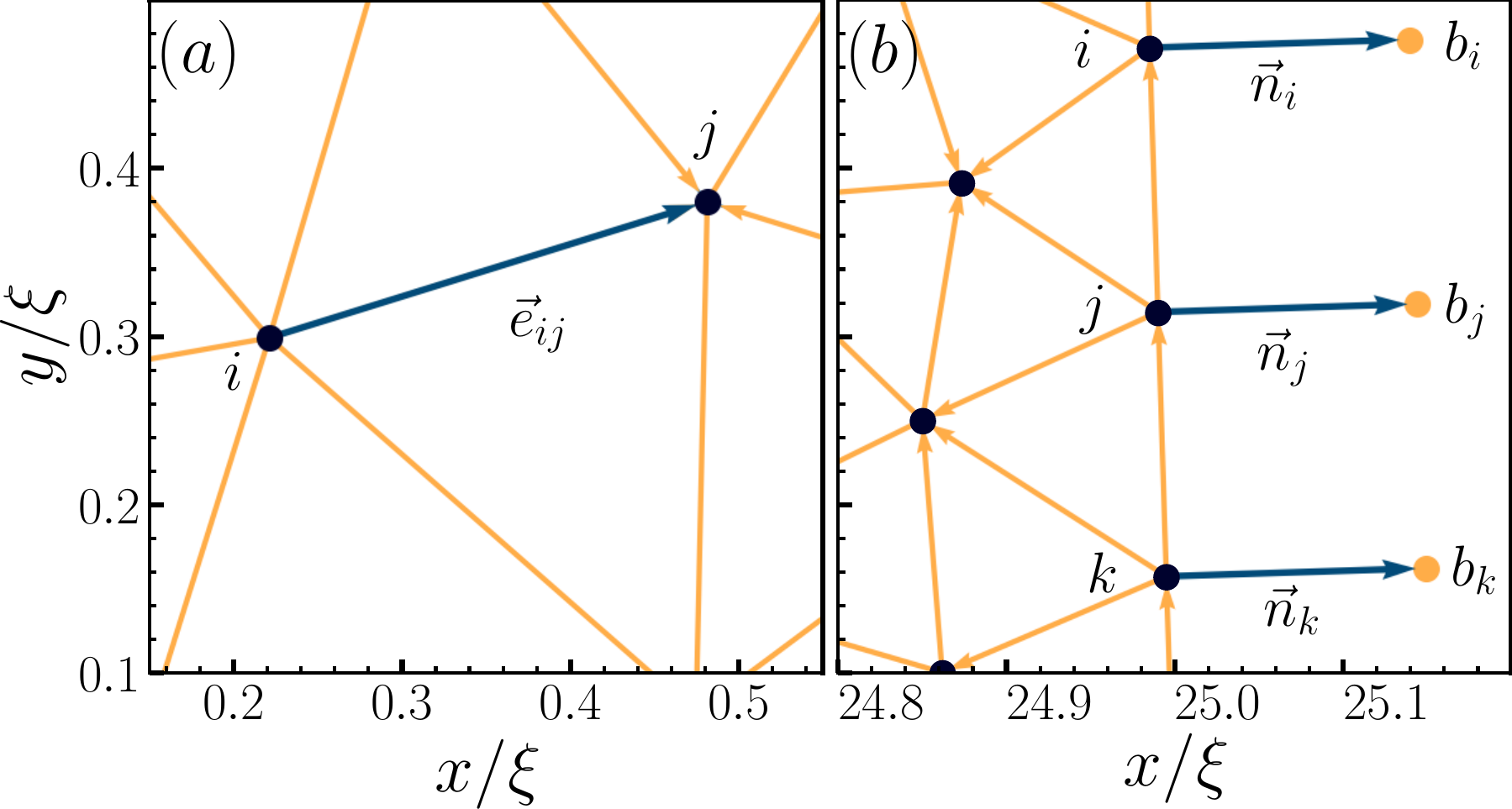}}
\end{minipage}
\caption{Schematic illustration of the  Delaunay mesh for the inner (a) and outer (b) parts of the film. The values of the order parameter $\psi$ and the supercurrent $J$  are calculated at each site in the mesh such as $i$ to obtain $\psi_{i}$ and $J_{i}$. Auxiliary boundary sites $b_{i}$ and $b_{j}$ are essential to numerically implement boundary conditions with sufficient accuracy. The actual amount of edges in the LSFG matrix exceeds that shown on the scheme. Nevertheless, the position of mesh's sites with respect to each other is determined by the Delaunay triangulation principles, and the maximum edge length limit is $0.5\xi$.}
\label{mesh}
\end{figure}

For numerical solution, we use the Delaunay mesh from the \textsc{pyTDGL} package \cite{pytdgl}. Figure \ref{mesh} shows the Delaunay mesh with normal vectors to the outer sides of the film, which are used to take into account the boundary conditions (\ref{cubs2}). As in  \textsc{pyTDGL}, distance is measured in units of $\xi$ - the superconducting coherence length. The dimensions of other variables are shown in the table:

\begin{center}
\begin{tabularx}{\columnwidth}{@{}>{\raggedright\arraybackslash}X 
                                >{\centering\arraybackslash}X@{}}
\toprule\toprule
\textbf{Quantity} & \textbf{Unit used in the simulations} \\ \midrule
Magnetic field               & $B_{c2}= \Phi_{0}/(2\pi\xi^{2})$ \\
Current density              & $0.4\,\xi B_{c2}/(\mu_{0}\lambda^{2})$ \\
Order-parameter amplitude    & $\sqrt{\alpha/\beta}$ \\
\bottomrule\bottomrule
\end{tabularx}
\end{center}
Here, $\Phi_{0} = h / 2e$ is the superconducting flux quantum, $\mu_0$ is the vacuum permeability, $\lambda$ is the London penetration depth, $\alpha, \beta$ are GL parameters mentioned in the introduction (\ref{Eq_GLeq}). The numerical simulation was implemented for all cases of GL($\eta, \gamma$) for radial film with $R = 25\xi$, $\xi = 0.5$, $\lambda = 4$. 

\section{Weighting coefficient matrix for LSFD}\label{sec:app_c}

LSFD is based on a two-dimensional Taylor series expansion for the value at a mesh point, where derivatives are treated as unknowns \cite{lsq}. However, the GL equation includes a covariant form of spatial derivatives. To compute these, it is possible to employ the Taylor series expansion while taking into account the spatial link variable $U(x,y)$  in discrete form\cite{link1}, \cite{link2}:
\begin{align}\label{cov_expan}
 &U_{ij}\psi_{j} = \sum_{n=0}^{\infty} \sum_{k=0}^{n}  (C_{n}^{k} / n!) \Delta x_{ij}^k \Delta y_{ij}^{n-k} \nonumber \\ & \qquad \qquad  \qquad \qquad (\partial_x - i\textbf{A}_x)_{i}^k
 (\partial_y - i\textbf{A}_y)_{i}^{n-k}\psi_{i}, \nonumber \\ &U_{ij} = \exp{(-i (\textbf{A}_{ij} \textbf{e}_{ij}))}, \\ &\textbf{e}_{ij} = \begin{pmatrix} \Delta x_{ij} \\\Delta  y_{ij} \end{pmatrix}, \; \textbf{A}_{ij} = \frac{B}{4} \begin{pmatrix} x_{i} + x_{j} \\ -(y_{i} + y_{j})\end{pmatrix}. \nonumber 
\end{align}

As shown in Figure \ref{mesh}, $i$ is the site on triangular mesh with known $\psi_{i}$, where we calculate covariant derivatives. $j$ is the neighboring site with known $\psi_{j}$, $\textbf{e}_{ij}$ is the edge between two sites: $\Delta x_{ij} = x_j - x_{i}$, $\Delta y_{ij} = y_j - y_{i}$. Figure \ref{mesh} does not illustrate the exact number of edges between the sites, as the actual number of edges is from 20 to 170. Increasing the number of edges is essential for the stable numerical solution of GL($\eta, \gamma$),   especially when the coefficient $\gamma$ is high.

We denote $\Delta \psi_{ij}^{c} = U_{ij}\psi_{j} - \psi_{i}, \, \partial_x^{c} = (\partial_x - i\textbf{A}_x)_{i}, \, \partial_y^{c} = (\partial_y - i\textbf{A}_y)_{i}$ . By truncating the Taylor series at N neighboring points, from the (\ref{cov_expan}) we obtained N equations with 14 unknown derivatives for 4th order accuracy:
\begin{align}
&\Delta\psi_{ij}^{c} = S_i \, d\psi_i^{c},  \\
&d\psi_i^{c} = [\partial_x^{c} \psi_i, \, \partial_y^{c} \psi_i, \dots,  \partial_{xxxy}^{c} \psi_i, \,\partial_{xyyy}^{c} \psi_i, \, \partial_{xxyy}^{c} \psi_i], \nonumber \\
&s_{ij} = \left[\Delta x_{ij}, \, \Delta y_{ij}, \dots,  \Delta x_{ij}^2\Delta y_{ij}^2 / 4 \right], \; j = 1 \dots N. \nonumber
\end{align}
In its current form, the system is inconsistent, hence it is required to apply least-square optimization with a number of neighboring sites $ N > 14$:
\begin{align}
d\psi_{i}^{c} = (S_{i}^{T}S_{i})^{-1} S_{i}^{T} \Delta \psi_{i}^{c}.
\end{align}

We also implemented the local scaling technique by scaling the local distance $d_{i} = max \Big( \sqrt{\Delta x_{ij}^2 + \Delta y_{ij}^2} \Big)$ in form  $ \; \bar x_{i} = \frac{\Delta x_{ij}}{d_{i}}, \; \bar y_{i} = \frac{\Delta y_{ij}}{d_{i}} , \, j = 1 \dots N$  to prevent the coefficient matrix $S_{i}$ from being ill-conditioned. The new form of the S matrix is:
\begin{align}\label{bigS}
\bar S_i  = 
\begin{bmatrix}
\Delta \bar x_{1} &\!\!\! \Delta \bar y_{1}  &\!\!\! \dots &\!\!\! \frac{\Delta \bar x_{1}^3\Delta \bar y_{1}}{6}  &\!\!\! \frac{\Delta \bar x_{1}\Delta \bar y_{1}^3}{6} &\!\!\! \frac{\Delta \bar x_{1}^2\Delta \bar y_{1}^2}{4}\!\!\!\\[1ex] 
\Delta \bar x_{2} &\!\!\! \Delta \bar y_{2}  &\!\!\! \dots &\!\!\! \frac{\Delta \bar x_{2}^3\Delta \bar y_{2}}{6}  &\!\!\! \frac{\Delta \bar x_{2}\Delta \bar y_{2}^3}{6} &\!\!\! \frac{\Delta \bar x_{2}^2\Delta \bar y_{2}^2}{4}\\[1ex] 
\vdots &\!\!\! \vdots &\!\!\! \vdots&\!\!\! \vdots&\!\!\!  \vdots&\!\!\! \vdots &\!\!\! \\[1ex]  
\Delta \bar x_{N} &\!\!\! \Delta \bar y_{N} &\!\!\!  \dots &\!\!\! \frac{\Delta \bar x_{N}^3\Delta \bar y_{N}}{6}  &\!\!\! \frac{\Delta \bar x_{N}\Delta \bar y_{N}^3}{6} &\!\!\! \frac{\Delta \bar x_{N}^2\Delta \bar y_{N}^2}{4}\!\!\!
\end{bmatrix}.
\end{align}
The matrix $D$ corresponding to the local scaling technique is:
\begin{align}
D_i  = 
\begin{bmatrix}
d_{i}^{-1} & 0  & \dots & 0 & 0 \\
0 & d_{i}^{-1}  & \dots & 0 & 0 \\
\vdots & \vdots  & \ddots & \vdots & \vdots \\
0 & 0  & \dots & d_{i}^{-4}  & 0 \\
0 & 0  & \dots & 0  &  d_{i}^{-4}
\end{bmatrix}.
\end{align}

In order to minimize the effect of distant points, we have also introduced a weight matrix $W_{i}$ with weights $ w_{ij} = \left(\sqrt{\Delta x_{ij}^2 + \Delta y_{ij}^2}\right)^{-1}, \, j = 1..N$:
\begin{align}
W_i  = 
\begin{bmatrix}
w_{i1} & 0  & \dots & 0 & 0 \\
0 & w_{i2} & \dots & 0 & 0 \\
\vdots & \vdots  & \ddots & \vdots & \vdots \\
0 & 0  & \dots &  w_{iN-1}  & 0 \\
0 & 0  & \dots & 0  &  w_{iN}
\end{bmatrix}.
\end{align}
To summarize, the covariant derivatives of $i$ point can be calculated by:
\begin{align}
d\psi_{i}^{c} = D_{i}(\bar S_{i}^{T} W \bar S_{i})^{-1} \bar S_{i}^{T} W \Delta \psi_{i}^{c}.
\end{align}
Computationally, LSFD is efficient because it calculates all the mixed derivatives that arise when $\gamma \neq 0$ for any vector $\textbf{x}_0$. It also handles nontrivial boundary conditions (\ref{cubs2})  naturally by modifying the last two equations at the boundary points (\ref{bigS}):
\begin{align}
&s_{i \, N-1} = [n_x,\, n_y, \,0, \dots,\, 0], \nonumber \\
&s_{i \, N} = [0,\, 0,\, 2 n_x,\, n_x, \, n_y, \dots ,0],  \end{align}
where $\bf{n} = (n_x, n_y)$ is the normal vector to film boundaries.
Additionally, it is essential to include a standard numerical implementation for boundary conditions in (\ref{cubs2}) in the following way:
\begin{align}
&\psi_{b}^{t} = \exp{(i(\textbf{A}, \textbf{n}))}\Big(\psi_i^{t-1} + 0.5\Delta n_{ix}^{2}(\partial_x - i\textbf{A}_x)^{2}\psi_{i}^{t-1} \nonumber  \\ 
& + 0.5\Delta n_{iy}^{2}(\partial_y - i\textbf{A}_y)^{2}\psi_{i}^{t-1} \\
& +\Delta n_{ix}\Delta n_{iy}(\partial_x - i\textbf{A}_x)(\partial_y - i\textbf{A}_y)\psi_{i}^{t-1}\Big), \nonumber 
\end{align}
where $\psi_{b}^{t}$ is the value of the order parameter along the normal direction. The absence of the first derivatives is explained by the first boundary condition from (\ref{cubs2}) for the projection of the covariant gradient along the normal. We use the value of $\psi_{i}^{t-1}$ and its second derivatives from the previous simulation time step to compute $\psi_{b}^{t}$, as shown in Figure \ref{mesh} site $i$ is the closest site to the auxiliary site $b_{i}$ along the direction of $\bf{n}$. Auxiliary boundary sites $b$ are included in the matrix (\ref{bigS}) along with the inner sites.

Finally, as in \textsc{pyTDGL} \cite{pytdgl}, values $\psi_{i}^{t+1}$ are calculated according to the (\ref{cubs1}) with the explicit Euler's method with an adaptive time step:
\begin{align}
&\psi_{i}^{t+1} = \psi_{i}^{t} + \frac{\Delta \tau^{t}}{u} F_{i}^{t}(\psi), \\
&F_{i}^{t} = \left(1+(\eta\mathbf{x_0})^2 - |\psi_{i}^{t}|^2\right)\psi_{i}^{t} +  \left((\partial_x - i\textbf{A}_x)^2\psi\right)_{i}^{t} + \nonumber \\  &\left((\partial_y -i\textbf{A}_y)^2\psi\right)_{i}^{t} - 2i\gamma \Big( \left((\partial_x - i\textbf{A}_x)^3\psi\right)_{i}^{t} \nonumber \\ &+ \left((\partial_x - i\textbf{A}_x)(\partial_y -i\textbf{A}_y)^2\psi_{i}^{t}\right) \Big), \nonumber
\end{align}
where $\Delta \tau^{t}$ is the time step in $t \rightarrow t+1$ simulation step.

\bibliography{lit}   

\begin{thebibliography}{54}%
\makeatletter
\providecommand \@ifxundefined [1]{%
 \@ifx{#1\undefined}
}%
\providecommand \@ifnum [1]{%
 \ifnum #1\expandafter \@firstoftwo
 \else \expandafter \@secondoftwo
 \fi
}%
\providecommand \@ifx [1]{%
 \ifx #1\expandafter \@firstoftwo
 \else \expandafter \@secondoftwo
 \fi
}%
\providecommand \natexlab [1]{#1}%
\providecommand \enquote  [1]{``#1''}%
\providecommand \bibnamefont  [1]{#1}%
\providecommand \bibfnamefont [1]{#1}%
\providecommand \citenamefont [1]{#1}%
\providecommand \href@noop [0]{\@secondoftwo}%
\providecommand \href [0]{\begingroup \@sanitize@url \@href}%
\providecommand \@href[1]{\@@startlink{#1}\@@href}%
\providecommand \@@href[1]{\endgroup#1\@@endlink}%
\providecommand \@sanitize@url [0]{\catcode `\\12\catcode `\$12\catcode
  `\&12\catcode `\#12\catcode `\^12\catcode `\_12\catcode `\%12\relax}%
\providecommand \@@startlink[1]{}%
\providecommand \@@endlink[0]{}%
\providecommand \url  [0]{\begingroup\@sanitize@url \@url }%
\providecommand \@url [1]{\endgroup\@href {#1}{\urlprefix }}%
\providecommand \urlprefix  [0]{URL }%
\providecommand \Eprint [0]{\href }%
\providecommand \doibase [0]{https://doi.org/}%
\providecommand \selectlanguage [0]{\@gobble}%
\providecommand \bibinfo  [0]{\@secondoftwo}%
\providecommand \bibfield  [0]{\@secondoftwo}%
\providecommand \translation [1]{[#1]}%
\providecommand \BibitemOpen [0]{}%
\providecommand \bibitemStop [0]{}%
\providecommand \bibitemNoStop [0]{.\EOS\space}%
\providecommand \EOS [0]{\spacefactor3000\relax}%
\providecommand \BibitemShut  [1]{\csname bibitem#1\endcsname}%
\let\auto@bib@innerbib\@empty
\bibitem [{\citenamefont {Buzdin}(2005)}]{Buzdin-RMP-05}%
  \BibitemOpen
  \bibfield  {author} {\bibinfo {author} {\bibfnamefont {A.~I.}\ \bibnamefont
  {Buzdin}},\ }\bibfield  {title} {\bibinfo {title} {Proximity effects in
  superconductor-ferromagnet heterostructures},\ }\href
  {https://doi.org/10.1103/RevModPhys.77.935} {\bibfield  {journal} {\bibinfo
  {journal} {Rev. Mod. Phys.}\ }\textbf {\bibinfo {volume} {77}},\ \bibinfo
  {pages} {935} (\bibinfo {year} {2005})}\BibitemShut {NoStop}%
\bibitem [{\citenamefont {Mel’nikov}\ \emph {et~al.}(2022)\citenamefont
  {Mel’nikov}, \citenamefont {Mironov}, \citenamefont {Samokhvalov},\ and\
  \citenamefont {Buzdin}}]{Mironov-UFN-22}%
  \BibitemOpen
  \bibfield  {author} {\bibinfo {author} {\bibfnamefont {A.~S.}\ \bibnamefont
  {Mel’nikov}}, \bibinfo {author} {\bibfnamefont {S.~V.}\ \bibnamefont
  {Mironov}}, \bibinfo {author} {\bibfnamefont {A.~V.}\ \bibnamefont
  {Samokhvalov}},\ and\ \bibinfo {author} {\bibfnamefont {A.~I.}\ \bibnamefont
  {Buzdin}},\ }\bibfield  {title} {\bibinfo {title} {Superconducting
  spintronics: state of the art and prospects},\ }\href
  {https://doi.org/10.3367/UFNe.2021.07.039020} {\bibfield  {journal} {\bibinfo
   {journal} {Phys. Usp.}\ }\textbf {\bibinfo {volume} {65}},\ \bibinfo {pages}
  {1248} (\bibinfo {year} {2022})}\BibitemShut {NoStop}%
\bibitem [{\citenamefont {Bergeret}\ \emph {et~al.}(2005)\citenamefont
  {Bergeret}, \citenamefont {Volkov},\ and\ \citenamefont
  {Efetov}}]{Bergeret-RMP-05}%
  \BibitemOpen
  \bibfield  {author} {\bibinfo {author} {\bibfnamefont {F.~S.}\ \bibnamefont
  {Bergeret}}, \bibinfo {author} {\bibfnamefont {A.~F.}\ \bibnamefont
  {Volkov}},\ and\ \bibinfo {author} {\bibfnamefont {K.~B.}\ \bibnamefont
  {Efetov}},\ }\bibfield  {title} {\bibinfo {title} {Odd triplet
  superconductivity and related phenomena in superconductor-ferromagnet
  structures},\ }\href {https://doi.org/10.1103/RevModPhys.77.1321} {\bibfield
  {journal} {\bibinfo  {journal} {Rev. Mod. Phys.}\ }\textbf {\bibinfo {volume}
  {77}},\ \bibinfo {pages} {1321} (\bibinfo {year} {2005})}\BibitemShut
  {NoStop}%
\bibitem [{\citenamefont {Alicea}(2012)}]{Alicea}%
  \BibitemOpen
  \bibfield  {author} {\bibinfo {author} {\bibfnamefont {J.}~\bibnamefont
  {Alicea}},\ }\bibfield  {title} {\bibinfo {title} {New directions in the
  pursuit of majorana fermions in solid-state systems},\ }\href
  {https://doi.org/10.1088/0034-4885/75/7/076501} {\bibfield  {journal}
  {\bibinfo  {journal} {Reports on Progress in Physics}\ }\textbf {\bibinfo
  {volume} {75}},\ \bibinfo {pages} {076501} (\bibinfo {year}
  {2012})}\BibitemShut {NoStop}%
\bibitem [{\citenamefont {Buzdin}(2008)}]{Buzdin_Phi}%
  \BibitemOpen
  \bibfield  {author} {\bibinfo {author} {\bibfnamefont {A.}~\bibnamefont
  {Buzdin}},\ }\bibfield  {title} {\bibinfo {title} {Direct coupling between
  magnetism and superconducting current in the josephson $\varphi_{0}$
  junction},\ }\href {https://doi.org/10.1103/PhysRevLett.101.107005}
  {\bibfield  {journal} {\bibinfo  {journal} {Physical Review Letters}\
  }\textbf {\bibinfo {volume} {101}},\ \bibinfo {pages} {107005} (\bibinfo
  {year} {2008})}\BibitemShut {NoStop}%
\bibitem [{\citenamefont {Krive}\ \emph {et~al.}(2004)\citenamefont {Krive},
  \citenamefont {Gorelik}, \citenamefont {Shekhter},\ and\ \citenamefont
  {Jonson}}]{Krive}%
  \BibitemOpen
  \bibfield  {author} {\bibinfo {author} {\bibfnamefont {I.~V.}\ \bibnamefont
  {Krive}}, \bibinfo {author} {\bibfnamefont {L.~Y.}\ \bibnamefont {Gorelik}},
  \bibinfo {author} {\bibfnamefont {R.~I.}\ \bibnamefont {Shekhter}},\ and\
  \bibinfo {author} {\bibfnamefont {M.}~\bibnamefont {Jonson}},\ }\bibfield
  {title} {\bibinfo {title} {Chiral symmetry breaking and the josephson current
  in a ballistic superconductor--quantum-wire--superconductor junction},\
  }\href {https://doi.org/10.1063/1.1739160} {\bibfield  {journal} {\bibinfo
  {journal} {Low Temperature Physics}\ }\textbf {\bibinfo {volume} {30}},\
  \bibinfo {pages} {398} (\bibinfo {year} {2004})}\BibitemShut {NoStop}%
\bibitem [{\citenamefont {Reynoso}\ \emph {et~al.}(2008)\citenamefont
  {Reynoso}, \citenamefont {Usaj}, \citenamefont {Balseiro}, \citenamefont
  {Feinberg},\ and\ \citenamefont {Avignon}}]{Reynoso}%
  \BibitemOpen
  \bibfield  {author} {\bibinfo {author} {\bibfnamefont {A.~A.}\ \bibnamefont
  {Reynoso}}, \bibinfo {author} {\bibfnamefont {G.}~\bibnamefont {Usaj}},
  \bibinfo {author} {\bibfnamefont {C.~A.}\ \bibnamefont {Balseiro}}, \bibinfo
  {author} {\bibfnamefont {D.}~\bibnamefont {Feinberg}},\ and\ \bibinfo
  {author} {\bibfnamefont {M.}~\bibnamefont {Avignon}},\ }\bibfield  {title}
  {\bibinfo {title} {Anomalous josephson current in junctions with
  spin-polarizing quantum point contacts},\ }\href
  {https://doi.org/10.1103/PhysRevLett.101.107001} {\bibfield  {journal}
  {\bibinfo  {journal} {Physical Review Letters}\ }\textbf {\bibinfo {volume}
  {101}},\ \bibinfo {pages} {107001} (\bibinfo {year} {2008})}\BibitemShut
  {NoStop}%
\bibitem [{\citenamefont {Szombati}\ \emph {et~al.}(2016)\citenamefont
  {Szombati}, \citenamefont {Nadj-Perge}, \citenamefont {Car}, \citenamefont
  {Plissard}, \citenamefont {Bakkers},\ and\ \citenamefont
  {Kouwenhoven}}]{Kouwenhoven}%
  \BibitemOpen
  \bibfield  {author} {\bibinfo {author} {\bibfnamefont {D.~B.}\ \bibnamefont
  {Szombati}}, \bibinfo {author} {\bibfnamefont {S.}~\bibnamefont
  {Nadj-Perge}}, \bibinfo {author} {\bibfnamefont {D.}~\bibnamefont {Car}},
  \bibinfo {author} {\bibfnamefont {S.~R.}\ \bibnamefont {Plissard}}, \bibinfo
  {author} {\bibfnamefont {E.~P. A.~M.}\ \bibnamefont {Bakkers}},\ and\
  \bibinfo {author} {\bibfnamefont {L.~P.}\ \bibnamefont {Kouwenhoven}},\
  }\bibfield  {title} {\bibinfo {title} {Josephson $\varphi_{0}$-junction in
  nanowire quantum dots},\ }\href {https://doi.org/10.1038/nphys3742}
  {\bibfield  {journal} {\bibinfo  {journal} {Nature Physics}\ }\textbf
  {\bibinfo {volume} {12}},\ \bibinfo {pages} {568} (\bibinfo {year}
  {2016})}\BibitemShut {NoStop}%
\bibitem [{\citenamefont {Mironov}\ and\ \citenamefont
  {Buzdin}(2017)}]{Mironov-PRL-17}%
  \BibitemOpen
  \bibfield  {author} {\bibinfo {author} {\bibfnamefont {S.}~\bibnamefont
  {Mironov}}\ and\ \bibinfo {author} {\bibfnamefont {A.}~\bibnamefont
  {Buzdin}},\ }\bibfield  {title} {\bibinfo {title} {Spontaneous currents in
  superconducting systems with strong spin–orbit coupling},\ }\href
  {https://doi.org/10.1103/PhysRevLett.118.077001} {\bibfield  {journal}
  {\bibinfo  {journal} {Physical Review Letters}\ }\textbf {\bibinfo {volume}
  {118}},\ \bibinfo {pages} {077001} (\bibinfo {year} {2017})}\BibitemShut
  {NoStop}%
\bibitem [{\citenamefont {Devizorova}\ \emph {et~al.}(2021)\citenamefont
  {Devizorova}, \citenamefont {Putilov}, \citenamefont {Chaykin}, \citenamefont
  {Mironov},\ and\ \citenamefont {Buzdin}}]{Devizorova-PRB-21}%
  \BibitemOpen
  \bibfield  {author} {\bibinfo {author} {\bibfnamefont {Z.}~\bibnamefont
  {Devizorova}}, \bibinfo {author} {\bibfnamefont {A.~V.}\ \bibnamefont
  {Putilov}}, \bibinfo {author} {\bibfnamefont {I.}~\bibnamefont {Chaykin}},
  \bibinfo {author} {\bibfnamefont {S.}~\bibnamefont {Mironov}},\ and\ \bibinfo
  {author} {\bibfnamefont {A.~I.}\ \bibnamefont {Buzdin}},\ }\bibfield  {title}
  {\bibinfo {title} {Phase transitions in superconductor/ferromagnet bilayer
  driven by spontaneous supercurrents},\ }\href
  {https://doi.org/10.1103/PhysRevB.103.064504} {\bibfield  {journal} {\bibinfo
   {journal} {Physical Review B}\ }\textbf {\bibinfo {volume} {103}},\ \bibinfo
  {pages} {064504} (\bibinfo {year} {2021})}\BibitemShut {NoStop}%
\bibitem [{\citenamefont {Krasnov}\ \emph {et~al.}(1997)\citenamefont
  {Krasnov}, \citenamefont {Oboznov},\ and\ \citenamefont
  {Pedersen}}]{Krasnov-PRB-97}%
  \BibitemOpen
  \bibfield  {author} {\bibinfo {author} {\bibfnamefont {V.~M.}\ \bibnamefont
  {Krasnov}}, \bibinfo {author} {\bibfnamefont {V.~A.}\ \bibnamefont
  {Oboznov}},\ and\ \bibinfo {author} {\bibfnamefont {N.~F.}\ \bibnamefont
  {Pedersen}},\ }\bibfield  {title} {\bibinfo {title} {Fluxon dynamics in long
  josephson junctions in the presence of a temperature gradient or spatial
  nonuniformity},\ }\href {https://doi.org/10.1103/PhysRevB.55.14486}
  {\bibfield  {journal} {\bibinfo  {journal} {Phys. Rev. B}\ }\textbf {\bibinfo
  {volume} {55}},\ \bibinfo {pages} {14486} (\bibinfo {year}
  {1997})}\BibitemShut {NoStop}%
\bibitem [{\citenamefont {Villegas}\ \emph {et~al.}(2003)\citenamefont
  {Villegas}, \citenamefont {Savel'ev}, \citenamefont {Nori}, \citenamefont
  {Gonzalez}, \citenamefont {Anguita}, \citenamefont {García},\ and\
  \citenamefont {Vicent}}]{Villegas-Sci-03}%
  \BibitemOpen
  \bibfield  {author} {\bibinfo {author} {\bibfnamefont {J.~E.}\ \bibnamefont
  {Villegas}}, \bibinfo {author} {\bibfnamefont {S.}~\bibnamefont {Savel'ev}},
  \bibinfo {author} {\bibfnamefont {F.}~\bibnamefont {Nori}}, \bibinfo {author}
  {\bibfnamefont {E.~M.}\ \bibnamefont {Gonzalez}}, \bibinfo {author}
  {\bibfnamefont {J.~V.}\ \bibnamefont {Anguita}}, \bibinfo {author}
  {\bibfnamefont {R.}~\bibnamefont {García}},\ and\ \bibinfo {author}
  {\bibfnamefont {J.~L.}\ \bibnamefont {Vicent}},\ }\bibfield  {title}
  {\bibinfo {title} {A superconducting reversible rectifier that controls the
  motion of magnetic flux quanta},\ }\href
  {https://doi.org/10.1126/science.1090390} {\bibfield  {journal} {\bibinfo
  {journal} {Science}\ }\textbf {\bibinfo {volume} {302}},\ \bibinfo {pages}
  {1188} (\bibinfo {year} {2003})}\BibitemShut {NoStop}%
\bibitem [{\citenamefont {Silaev}\ \emph {et~al.}(2014)\citenamefont {Silaev},
  \citenamefont {Aladyshkin}, \citenamefont {Silaeva},\ and\ \citenamefont
  {Aladyshkina}}]{Silaev-JPCM-14}%
  \BibitemOpen
  \bibfield  {author} {\bibinfo {author} {\bibfnamefont {M.~A.}\ \bibnamefont
  {Silaev}}, \bibinfo {author} {\bibfnamefont {A.~Y.}\ \bibnamefont
  {Aladyshkin}}, \bibinfo {author} {\bibfnamefont {M.~V.}\ \bibnamefont
  {Silaeva}},\ and\ \bibinfo {author} {\bibfnamefont {A.~S.}\ \bibnamefont
  {Aladyshkina}},\ }\bibfield  {title} {\bibinfo {title} {The diode effect
  induced by domain-wall superconductivity},\ }\href
  {https://doi.org/10.1088/0953-8984/26/9/095702} {\bibfield  {journal}
  {\bibinfo  {journal} {Journal of Physics: Condensed Matter}\ }\textbf
  {\bibinfo {volume} {26}},\ \bibinfo {pages} {095702} (\bibinfo {year}
  {2014})}\BibitemShut {NoStop}%
\bibitem [{\citenamefont {Lyu}\ \emph {et~al.}(2021)\citenamefont {Lyu},
  \citenamefont {Jiang}, \citenamefont {Wang}, \citenamefont {Xiao},
  \citenamefont {Dong}, \citenamefont {Chen}, \citenamefont {Milošević},
  \citenamefont {Wang}, \citenamefont {Divan}, \citenamefont {Pearson},
  \citenamefont {Wu}, \citenamefont {Peeters},\ and\ \citenamefont
  {Kwok}}]{Lyu-NatComm-21}%
  \BibitemOpen
  \bibfield  {author} {\bibinfo {author} {\bibfnamefont {Y.-Y.}\ \bibnamefont
  {Lyu}}, \bibinfo {author} {\bibfnamefont {J.}~\bibnamefont {Jiang}}, \bibinfo
  {author} {\bibfnamefont {Y.-L.}\ \bibnamefont {Wang}}, \bibinfo {author}
  {\bibfnamefont {Z.-L.}\ \bibnamefont {Xiao}}, \bibinfo {author}
  {\bibfnamefont {S.}~\bibnamefont {Dong}}, \bibinfo {author} {\bibfnamefont
  {Q.-H.}\ \bibnamefont {Chen}}, \bibinfo {author} {\bibfnamefont {M.~V.}\
  \bibnamefont {Milošević}}, \bibinfo {author} {\bibfnamefont
  {H.}~\bibnamefont {Wang}}, \bibinfo {author} {\bibfnamefont {R.}~\bibnamefont
  {Divan}}, \bibinfo {author} {\bibfnamefont {J.~E.}\ \bibnamefont {Pearson}},
  \bibinfo {author} {\bibfnamefont {P.}~\bibnamefont {Wu}}, \bibinfo {author}
  {\bibfnamefont {F.~M.}\ \bibnamefont {Peeters}},\ and\ \bibinfo {author}
  {\bibfnamefont {W.-K.}\ \bibnamefont {Kwok}},\ }\bibfield  {title} {\bibinfo
  {title} {Superconducting diode effect via conformal-mapped nanoholes},\
  }\href {https://doi.org/10.1038/s41467-021-23077-0} {\bibfield  {journal}
  {\bibinfo  {journal} {Nature Communications}\ }\textbf {\bibinfo {volume}
  {12}},\ \bibinfo {pages} {2703} (\bibinfo {year} {2021})}\BibitemShut
  {NoStop}%
\bibitem [{\citenamefont {Majer}\ \emph {et~al.}(2003)\citenamefont {Majer},
  \citenamefont {Peguiron}, \citenamefont {Grifoni}, \citenamefont {Tusveld},\
  and\ \citenamefont {Mooij}}]{Majer-PRL-03}%
  \BibitemOpen
  \bibfield  {author} {\bibinfo {author} {\bibfnamefont {J.~B.}\ \bibnamefont
  {Majer}}, \bibinfo {author} {\bibfnamefont {J.}~\bibnamefont {Peguiron}},
  \bibinfo {author} {\bibfnamefont {M.}~\bibnamefont {Grifoni}}, \bibinfo
  {author} {\bibfnamefont {M.}~\bibnamefont {Tusveld}},\ and\ \bibinfo {author}
  {\bibfnamefont {J.~E.}\ \bibnamefont {Mooij}},\ }\bibfield  {title} {\bibinfo
  {title} {Quantum ratchet effect for vortices},\ }\href
  {https://doi.org/10.1103/PhysRevLett.90.056802} {\bibfield  {journal}
  {\bibinfo  {journal} {Phys. Rev. Lett.}\ }\textbf {\bibinfo {volume} {90}},\
  \bibinfo {pages} {056802} (\bibinfo {year} {2003})}\BibitemShut {NoStop}%
\bibitem [{\citenamefont {Vodolazov}\ \emph {et~al.}(2005)\citenamefont
  {Vodolazov}, \citenamefont {Gribkov}, \citenamefont {Gusev}, \citenamefont
  {Klimov}, \citenamefont {Nozdrin}, \citenamefont {Rogov},\ and\ \citenamefont
  {Vdovichev}}]{Vodolazov-PRB-05}%
  \BibitemOpen
  \bibfield  {author} {\bibinfo {author} {\bibfnamefont {D.~Y.}\ \bibnamefont
  {Vodolazov}}, \bibinfo {author} {\bibfnamefont {B.~A.}\ \bibnamefont
  {Gribkov}}, \bibinfo {author} {\bibfnamefont {S.~A.}\ \bibnamefont {Gusev}},
  \bibinfo {author} {\bibfnamefont {A.~Y.}\ \bibnamefont {Klimov}}, \bibinfo
  {author} {\bibfnamefont {Y.~N.}\ \bibnamefont {Nozdrin}}, \bibinfo {author}
  {\bibfnamefont {V.~V.}\ \bibnamefont {Rogov}},\ and\ \bibinfo {author}
  {\bibfnamefont {S.~N.}\ \bibnamefont {Vdovichev}},\ }\bibfield  {title}
  {\bibinfo {title} {Considerable enhancement of the critical current in a
  superconducting film by a magnetized magnetic strip},\ }\href
  {https://doi.org/10.1103/PhysRevB.72.064509} {\bibfield  {journal} {\bibinfo
  {journal} {Phys. Rev. B}\ }\textbf {\bibinfo {volume} {72}},\ \bibinfo
  {pages} {064509} (\bibinfo {year} {2005})}\BibitemShut {NoStop}%
\bibitem [{\citenamefont {Souza~Silva}\ \emph {et~al.}(2006)\citenamefont
  {Souza~Silva}, \citenamefont {Van~de Vondel}, \citenamefont {Morelle},\ and\
  \citenamefont {Moshchalkov}}]{deSilva-NatLett-06}%
  \BibitemOpen
  \bibfield  {author} {\bibinfo {author} {\bibfnamefont {C.~C.~d.}\
  \bibnamefont {Souza~Silva}}, \bibinfo {author} {\bibfnamefont
  {J.}~\bibnamefont {Van~de Vondel}}, \bibinfo {author} {\bibfnamefont
  {M.}~\bibnamefont {Morelle}},\ and\ \bibinfo {author} {\bibfnamefont {V.~V.}\
  \bibnamefont {Moshchalkov}},\ }\bibfield  {title} {\bibinfo {title}
  {Controlled multiple reversals of a ratchet effect},\ }\href
  {https://doi.org/10.1038/nature04595} {\bibfield  {journal} {\bibinfo
  {journal} {Nature}\ }\textbf {\bibinfo {volume} {440}},\ \bibinfo {pages}
  {651} (\bibinfo {year} {2006})}\BibitemShut {NoStop}%
\bibitem [{\citenamefont {Suri}\ \emph {et~al.}(2022)\citenamefont {Suri},
  \citenamefont {Kamra}, \citenamefont {Meier}, \citenamefont {Kronseder},
  \citenamefont {Belzig}, \citenamefont {Back},\ and\ \citenamefont
  {Strunk}}]{Suri-APL-22}%
  \BibitemOpen
  \bibfield  {author} {\bibinfo {author} {\bibfnamefont {D.}~\bibnamefont
  {Suri}}, \bibinfo {author} {\bibfnamefont {A.}~\bibnamefont {Kamra}},
  \bibinfo {author} {\bibfnamefont {T.~N.~G.}\ \bibnamefont {Meier}}, \bibinfo
  {author} {\bibfnamefont {M.}~\bibnamefont {Kronseder}}, \bibinfo {author}
  {\bibfnamefont {W.}~\bibnamefont {Belzig}}, \bibinfo {author} {\bibfnamefont
  {C.~H.}\ \bibnamefont {Back}},\ and\ \bibinfo {author} {\bibfnamefont
  {C.}~\bibnamefont {Strunk}},\ }\bibfield  {title} {\bibinfo {title}
  {Non-reciprocity of vortex-limited critical current in conventional
  superconducting micro-bridges},\ }\href {https://doi.org/10.1063/5.0109753}
  {\bibfield  {journal} {\bibinfo  {journal} {Applied Physics Letters}\
  }\textbf {\bibinfo {volume} {121}},\ \bibinfo {pages} {102601} (\bibinfo
  {year} {2022})}\BibitemShut {NoStop}%
\bibitem [{\citenamefont {Ando}\ \emph {et~al.}(2020)\citenamefont {Ando},
  \citenamefont {Miyasaka}, \citenamefont {Li}, \citenamefont {Ishizuka},
  \citenamefont {Arakawa}, \citenamefont {Shiota}, \citenamefont {Moriyama},
  \citenamefont {Yanase},\ and\ \citenamefont {Ono}}]{Ando-Nature-21}%
  \BibitemOpen
  \bibfield  {author} {\bibinfo {author} {\bibfnamefont {F.}~\bibnamefont
  {Ando}}, \bibinfo {author} {\bibfnamefont {Y.}~\bibnamefont {Miyasaka}},
  \bibinfo {author} {\bibfnamefont {T.}~\bibnamefont {Li}}, \bibinfo {author}
  {\bibfnamefont {J.}~\bibnamefont {Ishizuka}}, \bibinfo {author}
  {\bibfnamefont {T.}~\bibnamefont {Arakawa}}, \bibinfo {author} {\bibfnamefont
  {Y.}~\bibnamefont {Shiota}}, \bibinfo {author} {\bibfnamefont
  {T.}~\bibnamefont {Moriyama}}, \bibinfo {author} {\bibfnamefont
  {Y.}~\bibnamefont {Yanase}},\ and\ \bibinfo {author} {\bibfnamefont
  {T.}~\bibnamefont {Ono}},\ }\bibfield  {title} {\bibinfo {title} {Observation
  of superconducting diode effect},\ }\href
  {https://doi.org/10.1038/s41586-020-2590-4} {\bibfield  {journal} {\bibinfo
  {journal} {Nature}\ }\textbf {\bibinfo {volume} {584}},\ \bibinfo {pages}
  {373} (\bibinfo {year} {2020})}\BibitemShut {NoStop}%
\bibitem [{\citenamefont {Bauriedl}\ \emph {et~al.}(2022)\citenamefont
  {Bauriedl}, \citenamefont {Bäuml}, \citenamefont {Fuchs},\ and\
  \citenamefont {Baumgartner}}]{Bauriedl-NatComm-22}%
  \BibitemOpen
  \bibfield  {author} {\bibinfo {author} {\bibfnamefont {L.}~\bibnamefont
  {Bauriedl}}, \bibinfo {author} {\bibfnamefont {C.}~\bibnamefont {Bäuml}},
  \bibinfo {author} {\bibfnamefont {L.}~\bibnamefont {Fuchs}},\ and\ \bibinfo
  {author} {\bibfnamefont {C.~e.~a.}\ \bibnamefont {Baumgartner}},\ }\bibfield
  {title} {\bibinfo {title} {Supercurrent diode effect and magnetochiral
  anisotropy in few-layer nbse$_2$},\ }\href
  {https://doi.org/10.1038/s41467-022-31954-5} {\bibfield  {journal} {\bibinfo
  {journal} {Nature Communications}\ }\textbf {\bibinfo {volume} {13}},\
  \bibinfo {pages} {4266} (\bibinfo {year} {2022})}\BibitemShut {NoStop}%
\bibitem [{\citenamefont {Baumgartner}\ \emph {et~al.}(2022)\citenamefont
  {Baumgartner}, \citenamefont {Fuchs}, \citenamefont {Costa}, \citenamefont
  {Reinhardt}, \citenamefont {Gronin}, \citenamefont {Gardner}, \citenamefont
  {Lindemann}, \citenamefont {Manfra}, \citenamefont {Faria~Junior},
  \citenamefont {Kochan}, \citenamefont {Fabian}, \citenamefont {Paradiso},\
  and\ \citenamefont {Strunk}}]{Baumgartner-NatNano-22}%
  \BibitemOpen
  \bibfield  {author} {\bibinfo {author} {\bibfnamefont {C.}~\bibnamefont
  {Baumgartner}}, \bibinfo {author} {\bibfnamefont {L.}~\bibnamefont {Fuchs}},
  \bibinfo {author} {\bibfnamefont {A.}~\bibnamefont {Costa}}, \bibinfo
  {author} {\bibfnamefont {S.}~\bibnamefont {Reinhardt}}, \bibinfo {author}
  {\bibfnamefont {S.}~\bibnamefont {Gronin}}, \bibinfo {author} {\bibfnamefont
  {G.~C.}\ \bibnamefont {Gardner}}, \bibinfo {author} {\bibfnamefont
  {T.}~\bibnamefont {Lindemann}}, \bibinfo {author} {\bibfnamefont {M.~J.}\
  \bibnamefont {Manfra}}, \bibinfo {author} {\bibfnamefont {P.~E.}\
  \bibnamefont {Faria~Junior}}, \bibinfo {author} {\bibfnamefont
  {D.}~\bibnamefont {Kochan}}, \bibinfo {author} {\bibfnamefont
  {J.}~\bibnamefont {Fabian}}, \bibinfo {author} {\bibfnamefont
  {N.}~\bibnamefont {Paradiso}},\ and\ \bibinfo {author} {\bibfnamefont
  {C.}~\bibnamefont {Strunk}},\ }\bibfield  {title} {\bibinfo {title}
  {Supercurrent rectification and magnetochiral effects in symmetric
  {Josephson} junctions},\ }\href {https://doi.org/10.1038/s41565-021-01009-9}
  {\bibfield  {journal} {\bibinfo  {journal} {Nature Nanotechnology}\ }\textbf
  {\bibinfo {volume} {17}},\ \bibinfo {pages} {39} (\bibinfo {year}
  {2022})}\BibitemShut {NoStop}%
\bibitem [{\citenamefont {Narita}\ \emph {et~al.}(2022)\citenamefont {Narita},
  \citenamefont {Ishizuka}, \citenamefont {Kawarazaki}, \citenamefont {Kan},
  \citenamefont {Shiota}, \citenamefont {Moriyama}, \citenamefont {Shimakawa},
  \citenamefont {Ognev}, \citenamefont {Samardak}, \citenamefont {Yanase},\
  and\ \citenamefont {Ono}}]{Narita-NatNano-22}%
  \BibitemOpen
  \bibfield  {author} {\bibinfo {author} {\bibfnamefont {H.}~\bibnamefont
  {Narita}}, \bibinfo {author} {\bibfnamefont {J.}~\bibnamefont {Ishizuka}},
  \bibinfo {author} {\bibfnamefont {R.}~\bibnamefont {Kawarazaki}}, \bibinfo
  {author} {\bibfnamefont {D.}~\bibnamefont {Kan}}, \bibinfo {author}
  {\bibfnamefont {Y.}~\bibnamefont {Shiota}}, \bibinfo {author} {\bibfnamefont
  {T.}~\bibnamefont {Moriyama}}, \bibinfo {author} {\bibfnamefont
  {Y.}~\bibnamefont {Shimakawa}}, \bibinfo {author} {\bibfnamefont {A.~V.}\
  \bibnamefont {Ognev}}, \bibinfo {author} {\bibfnamefont {A.~S.}\ \bibnamefont
  {Samardak}}, \bibinfo {author} {\bibfnamefont {Y.}~\bibnamefont {Yanase}},\
  and\ \bibinfo {author} {\bibfnamefont {T.}~\bibnamefont {Ono}},\ }\bibfield
  {title} {\bibinfo {title} {Field-free superconducting diode effect in
  noncentrosymmetric superconductor/ferromagnet multilayers},\ }\href
  {https://doi.org/10.1038/s41565-022-01159-4} {\bibfield  {journal} {\bibinfo
  {journal} {Nature Nanotechnology}\ }\textbf {\bibinfo {volume} {17}},\
  \bibinfo {pages} {823} (\bibinfo {year} {2022})}\BibitemShut {NoStop}%
\bibitem [{\citenamefont {Fominov}\ and\ \citenamefont
  {Mikhailov}(2022)}]{Fominov-PRB-22}%
  \BibitemOpen
  \bibfield  {author} {\bibinfo {author} {\bibfnamefont {Y.~V.}\ \bibnamefont
  {Fominov}}\ and\ \bibinfo {author} {\bibfnamefont {D.~S.}\ \bibnamefont
  {Mikhailov}},\ }\bibfield  {title} {\bibinfo {title} {Josephson diode with
  helical edge states},\ }\href {https://doi.org/10.1103/PhysRevB.106.134514}
  {\bibfield  {journal} {\bibinfo  {journal} {Physical Review B}\ }\textbf
  {\bibinfo {volume} {106}},\ \bibinfo {pages} {134514} (\bibinfo {year}
  {2022})}\BibitemShut {NoStop}%
\bibitem [{\citenamefont {Daido}\ \emph {et~al.}(2022)\citenamefont {Daido},
  \citenamefont {Ikeda},\ and\ \citenamefont {Yanase}}]{Daido-PRL-22}%
  \BibitemOpen
  \bibfield  {author} {\bibinfo {author} {\bibfnamefont {A.}~\bibnamefont
  {Daido}}, \bibinfo {author} {\bibfnamefont {Y.}~\bibnamefont {Ikeda}},\ and\
  \bibinfo {author} {\bibfnamefont {Y.}~\bibnamefont {Yanase}},\ }\bibfield
  {title} {\bibinfo {title} {Intrinsic superconducting diode effect},\ }\href
  {https://doi.org/10.1103/PhysRevLett.128.037001} {\bibfield  {journal}
  {\bibinfo  {journal} {Physical Review Letters}\ }\textbf {\bibinfo {volume}
  {128}},\ \bibinfo {pages} {037001} (\bibinfo {year} {2022})}\BibitemShut
  {NoStop}%
\bibitem [{\citenamefont {Ili\'c}\ and\ \citenamefont
  {Bergeret}(2022)}]{Ilic-PRL-22}%
  \BibitemOpen
  \bibfield  {author} {\bibinfo {author} {\bibfnamefont {S.}~\bibnamefont
  {Ili\'c}}\ and\ \bibinfo {author} {\bibfnamefont {F.~S.}\ \bibnamefont
  {Bergeret}},\ }\bibfield  {title} {\bibinfo {title} {Theory of the
  superconducting diode effect},\ }\href
  {https://doi.org/10.1103/PhysRevLett.128.177001} {\bibfield  {journal}
  {\bibinfo  {journal} {Physical Review Letters}\ }\textbf {\bibinfo {volume}
  {128}},\ \bibinfo {pages} {177001} (\bibinfo {year} {2022})}\BibitemShut
  {NoStop}%
\bibitem [{\citenamefont {Putilov}\ \emph {et~al.}(2024)\citenamefont
  {Putilov}, \citenamefont {Mironov},\ and\ \citenamefont
  {Buzdin}}]{Putilov-PRB-24}%
  \BibitemOpen
  \bibfield  {author} {\bibinfo {author} {\bibfnamefont {A.~V.}\ \bibnamefont
  {Putilov}}, \bibinfo {author} {\bibfnamefont {S.~V.}\ \bibnamefont
  {Mironov}},\ and\ \bibinfo {author} {\bibfnamefont {A.~I.}\ \bibnamefont
  {Buzdin}},\ }\bibfield  {title} {\bibinfo {title} {Nonreciprocal electron
  transport in finite-size superconductor/ferromagnet bilayers with strong
  spin-orbit coupling},\ }\href {https://doi.org/10.1103/PhysRevB.109.014510}
  {\bibfield  {journal} {\bibinfo  {journal} {Physical Review B}\ }\textbf
  {\bibinfo {volume} {109}},\ \bibinfo {pages} {014510} (\bibinfo {year}
  {2024})}\BibitemShut {NoStop}%
\bibitem [{\citenamefont {Kochan}\ \emph {et~al.}(2023)\citenamefont {Kochan},
  \citenamefont {Costa}, \citenamefont {Zhumagulov},\ and\ \citenamefont
  {{\v{Z}}uti{\'c}}}]{Kochan-Arxiv-2023}%
  \BibitemOpen
  \bibfield  {author} {\bibinfo {author} {\bibfnamefont {D.}~\bibnamefont
  {Kochan}}, \bibinfo {author} {\bibfnamefont {A.}~\bibnamefont {Costa}},
  \bibinfo {author} {\bibfnamefont {I.}~\bibnamefont {Zhumagulov}},\ and\
  \bibinfo {author} {\bibfnamefont {I.}~\bibnamefont {{\v{Z}}uti{\'c}}},\
  }\bibfield  {title} {\bibinfo {title} {Phenomenological theory of the
  supercurrent diode effect: The lifshitz invariant},\ }\href@noop {}
  {\bibfield  {journal} {\bibinfo  {journal} {arXiv preprint arXiv:2303.11975}\
  } (\bibinfo {year} {2023})}\BibitemShut {NoStop}%
\bibitem [{\citenamefont {Mineev}\ and\ \citenamefont
  {Samokhin}(1994)}]{Mineev-JETP-94}%
  \BibitemOpen
  \bibfield  {author} {\bibinfo {author} {\bibfnamefont {V.~P.}\ \bibnamefont
  {Mineev}}\ and\ \bibinfo {author} {\bibfnamefont {K.~V.}\ \bibnamefont
  {Samokhin}},\ }\bibfield  {title} {\bibinfo {title} {Helical phases in
  superconductors},\ }\href@noop {} {\bibfield  {journal} {\bibinfo  {journal}
  {Zh. Eksp. Teor. Fiz.}\ }\textbf {\bibinfo {volume} {105}},\ \bibinfo {pages}
  {747} (\bibinfo {year} {1994})},\ \bibinfo {note} {[Sov. Phys. JETP 78, 401
  (1994)]}\BibitemShut {NoStop}%
\bibitem [{\citenamefont {Edelstein}(1996)}]{Edelstein-JPCM-96}%
  \BibitemOpen
  \bibfield  {author} {\bibinfo {author} {\bibfnamefont {V.~M.}\ \bibnamefont
  {Edelstein}},\ }\bibfield  {title} {\bibinfo {title} {The ginzburg–landau
  equation for superconductors of polar symmetry},\ }\href
  {https://doi.org/10.1088/0953-8984/8/3/002} {\bibfield  {journal} {\bibinfo
  {journal} {Journal of Physics: Condensed Matter}\ }\textbf {\bibinfo {volume}
  {8}},\ \bibinfo {pages} {339} (\bibinfo {year} {1996})}\BibitemShut {NoStop}%
\bibitem [{\citenamefont {Kaur}\ \emph {et~al.}(2005)\citenamefont {Kaur},
  \citenamefont {Agterberg},\ and\ \citenamefont {Sigrist}}]{Kaur-PRL-05}%
  \BibitemOpen
  \bibfield  {author} {\bibinfo {author} {\bibfnamefont {R.~P.}\ \bibnamefont
  {Kaur}}, \bibinfo {author} {\bibfnamefont {D.~F.}\ \bibnamefont
  {Agterberg}},\ and\ \bibinfo {author} {\bibfnamefont {M.}~\bibnamefont
  {Sigrist}},\ }\bibfield  {title} {\bibinfo {title} {Helical vortex phase in
  non-centrosymmetric superconductors},\ }\href
  {https://doi.org/10.1103/PhysRevLett.94.137002} {\bibfield  {journal}
  {\bibinfo  {journal} {Physical Review Letters}\ }\textbf {\bibinfo {volume}
  {94}},\ \bibinfo {pages} {137002} (\bibinfo {year} {2005})}\BibitemShut
  {NoStop}%
\bibitem [{\citenamefont {He}\ \emph {et~al.}(2022)\citenamefont {He},
  \citenamefont {Tanaka},\ and\ \citenamefont {Nagaosa}}]{He-NJP-22}%
  \BibitemOpen
  \bibfield  {author} {\bibinfo {author} {\bibfnamefont {J.~J.}\ \bibnamefont
  {He}}, \bibinfo {author} {\bibfnamefont {Y.}~\bibnamefont {Tanaka}},\ and\
  \bibinfo {author} {\bibfnamefont {N.}~\bibnamefont {Nagaosa}},\ }\bibfield
  {title} {\bibinfo {title} {Topological superconducting diode effect in rashba
  superconductors},\ }\href {https://doi.org/10.1088/1367-2630/ac6a13}
  {\bibfield  {journal} {\bibinfo  {journal} {New Journal of Physics}\ }\textbf
  {\bibinfo {volume} {24}},\ \bibinfo {pages} {053014} (\bibinfo {year}
  {2022})}\BibitemShut {NoStop}%
\bibitem [{\citenamefont {Mironov}\ \emph {et~al.}(2024)\citenamefont
  {Mironov}, \citenamefont {Mel'nikov},\ and\ \citenamefont
  {Buzdin}}]{Mironov-PRB-24}%
  \BibitemOpen
  \bibfield  {author} {\bibinfo {author} {\bibfnamefont {S.~V.}\ \bibnamefont
  {Mironov}}, \bibinfo {author} {\bibfnamefont {A.~S.}\ \bibnamefont
  {Mel'nikov}},\ and\ \bibinfo {author} {\bibfnamefont {A.~I.}\ \bibnamefont
  {Buzdin}},\ }\bibfield  {title} {\bibinfo {title} {Photogalvanic phenomena in
  superconductors supporting intrinsic diode effect},\ }\href
  {https://doi.org/10.1103/PhysRevB.109.L220503} {\bibfield  {journal}
  {\bibinfo  {journal} {Phys. Rev. B}\ }\textbf {\bibinfo {volume} {109}},\
  \bibinfo {pages} {L220503} (\bibinfo {year} {2024})}\BibitemShut {NoStop}%
\bibitem [{\citenamefont {Plastovets}\ and\ \citenamefont
  {Buzdin}(2024)}]{Plastovets-PRB-24}%
  \BibitemOpen
  \bibfield  {author} {\bibinfo {author} {\bibfnamefont {V.}~\bibnamefont
  {Plastovets}}\ and\ \bibinfo {author} {\bibfnamefont {A.}~\bibnamefont
  {Buzdin}},\ }\bibfield  {title} {\bibinfo {title} {Magnetoelectric effect in
  the helical state of a superconductor/ferromagnet bilayer},\ }\href
  {https://doi.org/10.1103/PhysRevB.110.144521} {\bibfield  {journal} {\bibinfo
   {journal} {Phys. Rev. B}\ }\textbf {\bibinfo {volume} {110}},\ \bibinfo
  {pages} {144521} (\bibinfo {year} {2024})}\BibitemShut {NoStop}%
\bibitem [{\citenamefont {Fuchs}\ \emph {et~al.}(2022)\citenamefont {Fuchs},
  \citenamefont {Kochan}, \citenamefont {Schmidt}, \citenamefont {H\"uttner},
  \citenamefont {Baumgartner}, \citenamefont {Reinhardt}, \citenamefont
  {Gronin}, \citenamefont {Gardner}, \citenamefont {Lindemann}, \citenamefont
  {Manfra}, \citenamefont {Strunk},\ and\ \citenamefont
  {Paradiso}}]{Fuchs-PRX-22}%
  \BibitemOpen
  \bibfield  {author} {\bibinfo {author} {\bibfnamefont {L.}~\bibnamefont
  {Fuchs}}, \bibinfo {author} {\bibfnamefont {D.}~\bibnamefont {Kochan}},
  \bibinfo {author} {\bibfnamefont {J.}~\bibnamefont {Schmidt}}, \bibinfo
  {author} {\bibfnamefont {N.}~\bibnamefont {H\"uttner}}, \bibinfo {author}
  {\bibfnamefont {C.}~\bibnamefont {Baumgartner}}, \bibinfo {author}
  {\bibfnamefont {S.}~\bibnamefont {Reinhardt}}, \bibinfo {author}
  {\bibfnamefont {S.}~\bibnamefont {Gronin}}, \bibinfo {author} {\bibfnamefont
  {G.~C.}\ \bibnamefont {Gardner}}, \bibinfo {author} {\bibfnamefont
  {T.}~\bibnamefont {Lindemann}}, \bibinfo {author} {\bibfnamefont {M.~J.}\
  \bibnamefont {Manfra}}, \bibinfo {author} {\bibfnamefont {C.}~\bibnamefont
  {Strunk}},\ and\ \bibinfo {author} {\bibfnamefont {N.}~\bibnamefont
  {Paradiso}},\ }\bibfield  {title} {\bibinfo {title} {Anisotropic vortex
  squeezing in synthetic rashba superconductors: A manifestation of lifshitz
  invariants},\ }\href {https://doi.org/10.1103/PhysRevX.12.041020} {\bibfield
  {journal} {\bibinfo  {journal} {Phys. Rev. X}\ }\textbf {\bibinfo {volume}
  {12}},\ \bibinfo {pages} {041020} (\bibinfo {year} {2022})}\BibitemShut
  {NoStop}%
\bibitem [{\citenamefont {Agterberg}\ and\ \citenamefont
  {Kaur}(2007)}]{Agterberg-PRB-07}%
  \BibitemOpen
  \bibfield  {author} {\bibinfo {author} {\bibfnamefont {D.~F.}\ \bibnamefont
  {Agterberg}}\ and\ \bibinfo {author} {\bibfnamefont {R.~P.}\ \bibnamefont
  {Kaur}},\ }\bibfield  {title} {\bibinfo {title} {Magnetic-field-induced
  helical and stripe phases in rashba superconductors},\ }\href
  {https://doi.org/10.1103/PhysRevB.75.064511} {\bibfield  {journal} {\bibinfo
  {journal} {Phys. Rev. B}\ }\textbf {\bibinfo {volume} {75}},\ \bibinfo
  {pages} {064511} (\bibinfo {year} {2007})}\BibitemShut {NoStop}%
\bibitem [{\citenamefont {Lu}\ and\ \citenamefont {Yip}(2008)}]{Lu-PRB-08}%
  \BibitemOpen
  \bibfield  {author} {\bibinfo {author} {\bibfnamefont {C.-K.}\ \bibnamefont
  {Lu}}\ and\ \bibinfo {author} {\bibfnamefont {S.}~\bibnamefont {Yip}},\
  }\bibfield  {title} {\bibinfo {title} {Signature of superconducting states in
  cubic crystal without inversion symmetry},\ }\href
  {https://doi.org/10.1103/PhysRevB.77.054515} {\bibfield  {journal} {\bibinfo
  {journal} {Phys. Rev. B}\ }\textbf {\bibinfo {volume} {77}},\ \bibinfo
  {pages} {054515} (\bibinfo {year} {2008})}\BibitemShut {NoStop}%
\bibitem [{\citenamefont {Lu}\ and\ \citenamefont {Yip}(2009)}]{Lu-JLTP-09}%
  \BibitemOpen
  \bibfield  {author} {\bibinfo {author} {\bibfnamefont {C.-K.}\ \bibnamefont
  {Lu}}\ and\ \bibinfo {author} {\bibfnamefont {S.}~\bibnamefont {Yip}},\
  }\bibfield  {title} {\bibinfo {title} {Transverse {Magnetic} {Field}
  {Distribution} in the {Vortex} {State} of {Noncentrosymmetric}
  {Superconductor} {with~O~Symmetry}},\ }\href
  {https://doi.org/10.1007/s10909-009-9876-0} {\bibfield  {journal} {\bibinfo
  {journal} {Journal of Low Temperature Physics}\ }\textbf {\bibinfo {volume}
  {155}},\ \bibinfo {pages} {160} (\bibinfo {year} {2009})}\BibitemShut
  {NoStop}%
\bibitem [{\citenamefont {Yip}(2005)}]{Yip-JLTP-2005}%
  \BibitemOpen
  \bibfield  {author} {\bibinfo {author} {\bibfnamefont {S.~K.}\ \bibnamefont
  {Yip}},\ }\bibfield  {title} {\bibinfo {title} {Magnetic {Properties} of a
  {Superconductor} with no {Inversion} {Symmetry}},\ }\href
  {https://doi.org/10.1007/s10909-005-6012-7} {\bibfield  {journal} {\bibinfo
  {journal} {Journal of Low Temperature Physics}\ }\textbf {\bibinfo {volume}
  {140}},\ \bibinfo {pages} {67} (\bibinfo {year} {2005})}\BibitemShut
  {NoStop}%
\bibitem [{\citenamefont {Pearl}(1964)}]{Pearl-APL-64}%
  \BibitemOpen
  \bibfield  {author} {\bibinfo {author} {\bibfnamefont {J.}~\bibnamefont
  {Pearl}},\ }\bibfield  {title} {\bibinfo {title} {Current distribution in
  superconducting films carrying quantized fluxoids},\ }\href
  {https://doi.org/10.1063/1.1754056} {\bibfield  {journal} {\bibinfo
  {journal} {Applied Physics Letters}\ }\textbf {\bibinfo {volume} {5}},\
  \bibinfo {pages} {65} (\bibinfo {year} {1964})}\BibitemShut {NoStop}%
\bibitem [{\citenamefont {Buzdin}\ and\ \citenamefont
  {Brison}(1994)}]{Buzdin-PLA-94}%
  \BibitemOpen
  \bibfield  {author} {\bibinfo {author} {\bibfnamefont {A.~I.}\ \bibnamefont
  {Buzdin}}\ and\ \bibinfo {author} {\bibfnamefont {J.~P.}\ \bibnamefont
  {Brison}},\ }\bibfield  {title} {\bibinfo {title} {Vortex structures in small
  superconducting disks},\ }\href
  {https://doi.org/10.1016/0375-9601(94)90155-4} {\bibfield  {journal}
  {\bibinfo  {journal} {Physics Letters A}\ }\textbf {\bibinfo {volume}
  {196}},\ \bibinfo {pages} {267} (\bibinfo {year} {1994})}\BibitemShut
  {NoStop}%
\bibitem [{\citenamefont {Tinkham}(1996)}]{Tinkham-book-96}%
  \BibitemOpen
  \bibfield  {author} {\bibinfo {author} {\bibfnamefont {M.}~\bibnamefont
  {Tinkham}},\ }\href
  {http://www.loc.gov/catdir/description/mh022/95022378.html} {\emph {\bibinfo
  {title} {Introduction to superconductivity}}},\ \bibinfo {edition} {2nd}\
  ed.\ (\bibinfo  {publisher} {McGraw Hill},\ \bibinfo {address} {New York},\
  \bibinfo {year} {1996})\BibitemShut {NoStop}%
\bibitem [{\citenamefont {Gaggioli}\ \emph {et~al.}(2024)\citenamefont
  {Gaggioli}, \citenamefont {Blatter}, \citenamefont {Novoselov},\ and\
  \citenamefont {Geshkenbein}}]{Gaggioli-PRR-2024}%
  \BibitemOpen
  \bibfield  {author} {\bibinfo {author} {\bibfnamefont {F.}~\bibnamefont
  {Gaggioli}}, \bibinfo {author} {\bibfnamefont {G.}~\bibnamefont {Blatter}},
  \bibinfo {author} {\bibfnamefont {K.~S.}\ \bibnamefont {Novoselov}},\ and\
  \bibinfo {author} {\bibfnamefont {V.~B.}\ \bibnamefont {Geshkenbein}},\
  }\bibfield  {title} {\bibinfo {title} {Superconductivity in atomically thin
  films: Two-dimensional critical state model},\ }\href
  {https://doi.org/10.1103/PhysRevResearch.6.023190} {\bibfield  {journal}
  {\bibinfo  {journal} {Phys. Rev. B}\ }\textbf {\bibinfo {volume} {6}},\
  \bibinfo {pages} {023190} (\bibinfo {year} {2024})}\BibitemShut {NoStop}%
\bibitem [{\citenamefont {Blatter}\ \emph {et~al.}(1994)\citenamefont
  {Blatter}, \citenamefont {Feigel'man}, \citenamefont {Geshkenbein},
  \citenamefont {Larkin},\ and\ \citenamefont {Vinokur}}]{Blatter-RMP-1994}%
  \BibitemOpen
  \bibfield  {author} {\bibinfo {author} {\bibfnamefont {G.}~\bibnamefont
  {Blatter}}, \bibinfo {author} {\bibfnamefont {M.~V.}\ \bibnamefont
  {Feigel'man}}, \bibinfo {author} {\bibfnamefont {V.~B.}\ \bibnamefont
  {Geshkenbein}}, \bibinfo {author} {\bibfnamefont {A.~I.}\ \bibnamefont
  {Larkin}},\ and\ \bibinfo {author} {\bibfnamefont {V.~M.}\ \bibnamefont
  {Vinokur}},\ }\bibfield  {title} {\bibinfo {title} {Vortices in
  high-temperature superconductors},\ }\href
  {https://doi.org/10.1103/RevModPhys.66.1125} {\bibfield  {journal} {\bibinfo
  {journal} {Rev. Mod. Phys.}\ }\textbf {\bibinfo {volume} {66}},\ \bibinfo
  {pages} {1125} (\bibinfo {year} {1994})}\BibitemShut {NoStop}%
\bibitem [{\citenamefont {Gor'kov}\ and\ \citenamefont
  {Kopnin}(1975)}]{Gorkov-UFN-1975}%
  \BibitemOpen
  \bibfield  {author} {\bibinfo {author} {\bibfnamefont {L.~P.}\ \bibnamefont
  {Gor'kov}}\ and\ \bibinfo {author} {\bibfnamefont {N.~B.}\ \bibnamefont
  {Kopnin}},\ }\bibfield  {title} {\bibinfo {title} {Vortex motion and
  resistivity of type-ii superconductors ina magnetic field},\ }\href@noop {}
  {\bibfield  {journal} {\bibinfo  {journal} {Soviet Physics Uspekhi}\ }\textbf
  {\bibinfo {volume} {18}},\ \bibinfo {pages} {496} (\bibinfo {year}
  {1975})}\BibitemShut {NoStop}%
\bibitem [{\citenamefont {Kopnin}(2001)}]{Kopnin-book-2001}%
  \BibitemOpen
  \bibfield  {author} {\bibinfo {author} {\bibfnamefont {N.}~\bibnamefont
  {Kopnin}},\ }\href {https://books.google.ru/books?id=BHVdDgAAQBAJ} {\emph
  {\bibinfo {title} {Theory of Nonequilibrium Superconductivity}}},\
  International Series of Monographs on Physics\ (\bibinfo  {publisher}
  {Clarendon Press},\ \bibinfo {year} {2001})\BibitemShut {NoStop}%
\bibitem [{\citenamefont {Genkin}\ and\ \citenamefont
  {Mel'nikov}(1989)}]{GenkinJETP}%
  \BibitemOpen
  \bibfield  {author} {\bibinfo {author} {\bibfnamefont {V.~M.}\ \bibnamefont
  {Genkin}}\ and\ \bibinfo {author} {\bibfnamefont {A.~S.}\ \bibnamefont
  {Mel'nikov}},\ }\bibfield  {title} {\bibinfo {title} {Motion of abrikosov
  vortices in anisotropic superconductors},\ }\href@noop {} {\bibfield
  {journal} {\bibinfo  {journal} {Zh. Eksp. Teor. Fiz.}\ }\textbf {\bibinfo
  {volume} {95}},\ \bibinfo {pages} {2170} (\bibinfo {year} {1989})},\ \bibinfo
  {note} {[Sov. Phys. JETP 68, 1254 (1989)]}\BibitemShut {NoStop}%
\bibitem [{\citenamefont {Hao}\ and\ \citenamefont
  {Clem}(1991)}]{Hao-IEEE-1991}%
  \BibitemOpen
  \bibfield  {author} {\bibinfo {author} {\bibfnamefont {Z.}~\bibnamefont
  {Hao}}\ and\ \bibinfo {author} {\bibfnamefont {J.}~\bibnamefont {Clem}},\
  }\bibfield  {title} {\bibinfo {title} {Viscous flux motion in anisotropic
  type-ii superconductors in low fields},\ }\href@noop {} {\bibfield  {journal}
  {\bibinfo  {journal} {IEEE transactions on magnetics}\ }\textbf {\bibinfo
  {volume} {27}},\ \bibinfo {pages} {1086} (\bibinfo {year}
  {1991})}\BibitemShut {NoStop}%
\bibitem [{\citenamefont {Sadovskyy}\ \emph {et~al.}(2015)\citenamefont
  {Sadovskyy}, \citenamefont {Koshelev}, \citenamefont {Phillips},
  \citenamefont {Karpeyev},\ and\ \citenamefont {Glatz}}]{Sadovskyy-JCP-2015}%
  \BibitemOpen
  \bibfield  {author} {\bibinfo {author} {\bibfnamefont {I.}~\bibnamefont
  {Sadovskyy}}, \bibinfo {author} {\bibfnamefont {A.}~\bibnamefont {Koshelev}},
  \bibinfo {author} {\bibfnamefont {C.}~\bibnamefont {Phillips}}, \bibinfo
  {author} {\bibfnamefont {D.}~\bibnamefont {Karpeyev}},\ and\ \bibinfo
  {author} {\bibfnamefont {A.}~\bibnamefont {Glatz}},\ }\bibfield  {title}
  {\bibinfo {title} {Stable large-scale solver for ginzburg–landau equations
  for superconductors},\ }\href
  {https://doi.org/https://doi.org/10.1016/j.jcp.2015.04.002} {\bibfield
  {journal} {\bibinfo  {journal} {Journal of Computational Physics}\ }\textbf
  {\bibinfo {volume} {294}},\ \bibinfo {pages} {639} (\bibinfo {year}
  {2015})}\BibitemShut {NoStop}%
\bibitem [{\citenamefont {Kato}\ \emph {et~al.}(1993)\citenamefont {Kato},
  \citenamefont {Enomoto},\ and\ \citenamefont {Maekawa}}]{Kato-PRB-1993}%
  \BibitemOpen
  \bibfield  {author} {\bibinfo {author} {\bibfnamefont {R.}~\bibnamefont
  {Kato}}, \bibinfo {author} {\bibfnamefont {Y.}~\bibnamefont {Enomoto}},\ and\
  \bibinfo {author} {\bibfnamefont {S.}~\bibnamefont {Maekawa}},\ }\bibfield
  {title} {\bibinfo {title} {Effects of the surface boundary on the
  magnetization process in type-ii superconductors},\ }\href@noop {} {\bibfield
   {journal} {\bibinfo  {journal} {Physical Review B}\ }\textbf {\bibinfo
  {volume} {47}},\ \bibinfo {pages} {8016} (\bibinfo {year}
  {1993})}\BibitemShut {NoStop}%
\bibitem [{\citenamefont {Kopasov}\ \emph {et~al.}(2023)\citenamefont
  {Kopasov}, \citenamefont {Tsar'kov},\ and\ \citenamefont
  {Mel'nikov}}]{Kopasov-PRB-2023}%
  \BibitemOpen
  \bibfield  {author} {\bibinfo {author} {\bibfnamefont {A.~A.}\ \bibnamefont
  {Kopasov}}, \bibinfo {author} {\bibfnamefont {I.~M.}\ \bibnamefont
  {Tsar'kov}},\ and\ \bibinfo {author} {\bibfnamefont {A.~S.}\ \bibnamefont
  {Mel'nikov}},\ }\bibfield  {title} {\bibinfo {title} {Disorder-induced
  trapping and antitrapping of vortices in type-ii superconductors},\ }\href
  {https://doi.org/10.1103/PhysRevB.107.174505} {\bibfield  {journal} {\bibinfo
   {journal} {Phys. Rev. B}\ }\textbf {\bibinfo {volume} {107}},\ \bibinfo
  {pages} {174505} (\bibinfo {year} {2023})}\BibitemShut {NoStop}%
\bibitem [{\citenamefont {Bishop-Van~Horn}(2023)}]{pytdgl}%
  \BibitemOpen
  \bibfield  {author} {\bibinfo {author} {\bibfnamefont {L.}~\bibnamefont
  {Bishop-Van~Horn}},\ }\bibfield  {title} {\bibinfo {title} {pytdgl: A
  flexible finite-element solver for time-dependent ginzburg–landau
  equations},\ }\href {https://doi.org/10.1016/j.cpc.2023.108799} {\bibfield
  {journal} {\bibinfo  {journal} {Computer Physics Communications}\ }\textbf
  {\bibinfo {volume} {291}},\ \bibinfo {pages} {108799} (\bibinfo {year}
  {2023})}\BibitemShut {NoStop}%
\bibitem [{\citenamefont {Liu}\ \emph {et~al.}(2020)\citenamefont {Liu},
  \citenamefont {Shu}, \citenamefont {Zhang},\ and\ \citenamefont
  {Yang}}]{lsq}%
  \BibitemOpen
  \bibfield  {author} {\bibinfo {author} {\bibfnamefont {Y.~Y.}\ \bibnamefont
  {Liu}}, \bibinfo {author} {\bibfnamefont {C.}~\bibnamefont {Shu}}, \bibinfo
  {author} {\bibfnamefont {H.~W.}\ \bibnamefont {Zhang}},\ and\ \bibinfo
  {author} {\bibfnamefont {L.~M.}\ \bibnamefont {Yang}},\ }\bibfield  {title}
  {\bibinfo {title} {High-order least-squares finite-difference schemes on
  unstructured meshes},\ }\href {https://doi.org/10.1016/j.jcp.2019.109019}
  {\bibfield  {journal} {\bibinfo  {journal} {Journal of Computational
  Physics}\ }\textbf {\bibinfo {volume} {401}},\ \bibinfo {pages} {109019}
  (\bibinfo {year} {2020})}\BibitemShut {NoStop}%
\bibitem [{\citenamefont {Gropp}\ \emph {et~al.}(1996)\citenamefont {Gropp},
  \citenamefont {Kaper}, \citenamefont {Leaf}, \citenamefont {Levine},
  \citenamefont {Palumbo},\ and\ \citenamefont {Vinokur}}]{link1}%
  \BibitemOpen
  \bibfield  {author} {\bibinfo {author} {\bibfnamefont {W.~D.}\ \bibnamefont
  {Gropp}}, \bibinfo {author} {\bibfnamefont {H.~G.}\ \bibnamefont {Kaper}},
  \bibinfo {author} {\bibfnamefont {G.~K.}\ \bibnamefont {Leaf}}, \bibinfo
  {author} {\bibfnamefont {D.~M.}\ \bibnamefont {Levine}}, \bibinfo {author}
  {\bibfnamefont {M.}~\bibnamefont {Palumbo}},\ and\ \bibinfo {author}
  {\bibfnamefont {V.~M.}\ \bibnamefont {Vinokur}},\ }\bibfield  {title}
  {\bibinfo {title} {Numerical simulation of vortex dynamics in type-ii
  superconductors},\ }\href {https://doi.org/10.1006/jcph.1996.0115} {\bibfield
   {journal} {\bibinfo  {journal} {Journal of Computational Physics}\ }\textbf
  {\bibinfo {volume} {123}},\ \bibinfo {pages} {254} (\bibinfo {year}
  {1996})}\BibitemShut {NoStop}%
\bibitem [{\citenamefont {Du}\ \emph {et~al.}(1998)\citenamefont {Du},
  \citenamefont {Nicolaides},\ and\ \citenamefont {Wu}}]{link2}%
  \BibitemOpen
  \bibfield  {author} {\bibinfo {author} {\bibfnamefont {Q.}~\bibnamefont
  {Du}}, \bibinfo {author} {\bibfnamefont {R.~A.}\ \bibnamefont {Nicolaides}},\
  and\ \bibinfo {author} {\bibfnamefont {X.}~\bibnamefont {Wu}},\ }\bibfield
  {title} {\bibinfo {title} {Analysis of finite difference schemes on irregular
  staggered grids},\ }\href {https://doi.org/10.1137/S0036142996291524}
  {\bibfield  {journal} {\bibinfo  {journal} {SIAM Journal on Numerical
  Analysis}\ }\textbf {\bibinfo {volume} {35}},\ \bibinfo {pages} {1044}
  (\bibinfo {year} {1998})}\BibitemShut {NoStop}%
\end{thebibliography}%
\end{document}